# The *Swift* Ultra-Violet/Optical Telescope

Peter W. A. Roming[*a], Thomas E. Kennedy[b], Keith O. Mason[b], John A. Nousek[a], Lindy Ahr[c],
Richard E. Bingham[d], Patrick S. Broos[a], Mary J. Carter[b], Barry K. Hancock[b],
Howard E. Huckle[b], S. D. Hunsberger[a], Hajime Kawakami[b], Ronnie Killough[c],
T. Scott Koch[a], Michael K. McLelland[c], Kelly Smith[c], Philip J. Smith[b],
Juan Carlos Soto[†e], Patricia T. Boyd[f], Alice A. Breeveld[b], Stephen T. Holland[f],
Mariya Ivanushkina[a], Michael S. Pryzby[g], Martin D. Still[f], Joseph Stock[g]

[a]Department of Astronomy & Astrophysics, Pennsylvania State University,
525 Davey Lab, University Park, PA 16802, USA

[b]Mullard Space Sciences Laboratory, University College London,
Holmbury St. Mary, Dorking, Surrey RH5 6NT, UK

[c]Southwest Research Institute,
6220 Culebra Rd, San Antonio, TX 78228, USA

[d]Optical Science Laboratory, University College London,
Gower St, London WC1E 6BT, UK

[e]Starsys Research Corporation,
4909 Nautilus Court N., Boulder, CO 80301, USA

[f]NASA/Goddard Space Flight Center,
Code 660.1, Greenbelt, MD 20771, USA

[g]Swales Aerospace,
5050 Powder Mill Rd, Beltsville, MD 20705, USA

**ABSTRACT**

The UV/Optical Telescope (UVOT) is one of three instruments flying aboard the *Swift* Gamma-ray Observatory. It is designed to capture the early (~1 minute) UV and optical photons from the afterglow of gamma-ray bursts in the 170-600 nm band as well as long term observations of these afterglows. This is accomplished through the use of UV and optical broadband filters and grisms. The UVOT has a modified Ritchey-Chrétien design with micro-channel plate intensified charged-coupled device detectors that record the arrival time of individual photons and provide sub-arcsecond positioning of sources. We discuss some of the science to be pursued by the UVOT and the overall design of the instrument.

## 1. INTRODUCTION

Since the initial discovery of gamma-ray bursts (GRBs; Klebesadel, Strong, & Olson 1973), slow progress has been made in understanding their nature due to their rapid decay in brightness. The burst phenomenon is typically very short-lived – on the order of tens of seconds for long bursts – and therefore very difficult to observe. It has become clear that the energies associated with these bursts are many orders of magnitude higher than supernova explosions (Mészáros & Rees 1992); they are the most energetic phenomena known in the Universe since the Big Bang (~$10^{53}$ ergs isotropic-equivalent).

---

[*] Correspondence: Email: roming@astro.psu.edu; Telephone: (814) 865-7745; Fax: (814) 865-6854
[†] Current Address: Ball Aerospace, 1600 Commerce St, Boulder, CO 80301, USA

In 1997, Mészáros & Rees suggested that x-ray, optical, and radio afterglows should be present after the initial GRB, thus allowing time to study these objects. Less than a month later, the first afterglow, from GRB 970228, was detected in the x-ray and optical bands (Costa *et al*. 1997). Later the same year, GRB 970508 provided evidence that GRBs are a cosmological phenomena when a redshift of z=0.835 was obtained for a galaxy coincident with the afterglow using the Mg II and Fe II absorption lines (Heise *et al*. 1997; Metzger *et al*. 1997). A radio afterglow was also discovered coincident with the same burst (Frail *et al*. 1997). A couple of years later, the first optical observation of the GRB prompt emission was achieved with the Robotic Optical Transient Search Experiment (ROTSE; Akerlof 1999).

Proposed progenitors of GRBs included neutron stars in our own galaxy (Higdon & Lingenfelter 1990), neutron star-neutron star and neutron star-black hole mergers (Goodman 1986; Paczyński 1986; Narayan, Paczyński, & Piran 1992), and "failed" supernovae (Woosley 1993) or hypernovae (Paczyński 1998). GRB 990123 was shown to be offset from the nucleus of the host galaxy and therefore provided evidence that the death and/or merger of massive stars were the origin of GRBs (Bloom 1999).

These and other studies began to unlock the secrets of GRBs. However, despite the rapid progress made in recent years in our understanding of GRBs, many questions still remain. What are the origins of GRBs? What can GRBs reveal about the early Universe? How many different classes and sub-classes of bursts exist? How does the blast wave develop and interact with its environment (Gehrels *et al*. 2004)?

The *Swift* Medium-class Explorer (MIDEX) mission is a rapidly slewing, multi-wavelength observatory designed to observe GRBs and their afterglows and to answer these questions. The observatory includes three telescopes: the Burst Alert Telescope (BAT; Barthelmy 2004), the X-Ray Telescope (XRT; Burrows *et al*. 2004), and the Ultra-Violet/Optical Telescope (UVOT). The wavelength coverage provided by these three instruments is 0.2-150 keV and 170-600 nm. Although other GRB missions, such as the *Compton Gamma Ray Observatory* (CGRO; 30 keV-30 GeV), *BeppoSAX* (0.1-200 keV), the *High Energy Transient Explorer-2* (*HETE-2*; 1-500 keV), and the *International Gamma-Ray Astrophysics Laboratory* (Integral; 3 keV-10 MeV and 500-850 nm), have covered the x-ray and gamma-ray regime that Swift covers, Swift is the first mission to cover the near-UV part of the spectrum. Overviews of the *Swift* mission are provided elsewhere (Gehrels *et al*. 2004; Nousek 2004).

In this paper, we discuss the to-be-flown UVOT. Discussions of earlier UVOT designs have been made elsewhere (Roming *et al*. 2000; Roming *et al*. 2004). In Section 2, we present an overview of the instrument and then describe the UVOT and its subsystems. In Sections 3 and 4, we furnish a description of the UVOT's observing scenarios and the instrument produced data products. In Sections 5 and 6, we provide a discussion of the instrument calibration and the science data products. Finally, in Sections 7 and 8, we outline some of the science goals of the UVOT and provide concluding remarks.

## 2. OVERVIEW OF UVOT

The UVOT has a modified Ritchey-Chrétien optical configuration with a 30 cm primary mirror and an f-number of 12.7. Light from the telescope is directed by a 45 degree mirror to one of two redundant detectors each of which lies behind an identical 11-position filter wheel. The filter wheel, detector, and associated processing electronics constitute two independent chains, one of which is held in cold redundancy on orbit. The filter wheels accommodate UV/optical grisms and broadband color filters, a 4x magnifier; a clear "white-light" filter, and a blocking filter. The wavelength range is 170-600 nm. The detectors are micro-channel plate intensified charged-coupled devices (MICs). These MICs image the incoming photons and operate in a photon counting mode. In the low sky background conditions afforded from Earth orbit, the UVOT will have a comparable imaging sensitivity to that of a 4 m ground-based telescope, able to detect a $24^{th}$ magnitude B-star in 1000s using the white-light filter. A brief synopsis of the UVOT's characteristics can be found in Table 1. The UVOT is mounted on the observatory optical bench with the BAT and the XRT as illustrated in Figure 1 and is co-aligned with the XRT.

The UVOT is made up of 5 functional units (see Figure 2): the telescope module (TM) consisting of the UV/optical telescope, a beam steering mirror, and the two redundant filter wheel mechanisms, photon counting detectors, power supplies, and electronics; two digital electronics modules (DEMs), each one housing a data processing unit (DPU), an instrument control unit (ICU), and power supplies for the DPU and ICU; and two interconnecting harness units to connect the TM to the two DEMs. The TM and ICU possess a strong legacy from the Optical/UV Monitor Telescope (OM) on ESA's XMM-Newton mission (Mason *et al.* 2001). Although the science goals of the OM are different than the UVOT, the OM TM design complements the objectives of the *Swift* mission because of its wavelength coverage from 170-600 nm and because of its broad sensitivity range. The DEM chassis is based on the SC-9 product (see http://www.swri.org/3pubs/ird1999/15909399.htm) which was used on the first MIDEX mission - the Imager for Magnetopause-to-Aurora Global Exploration (IMAGE; Gibson *et al.* 2000).

| Telescope | Modified Ritchey-Chrétien |
|---|---|
| Aperture | 30 cm diameter |
| f-number | 12.7 |
| Filters | 11 |
| Wavelength Range | 170-600 nm |
| Detector | MCP Intensified CCD |
| Detector Operation | Photon Counting |
| Sensitivity | $m_B$=24.0 in white light in 1000s |
| Field of View | 17 x 17 arcmin$^2$ |
| Detection Element | 256 x 256 pixels |
| Sampling Element | 2048 x 2048 after centroiding |
| Telescope PSF | 0.9 arcsec FWHM @ 350nm |
| Pixel Scale | 0.5 arcsec |

**Table 1. UVOT Characteristics**

Some changes were incorporated into the UVOT based on experience in operating XMM-OM in orbit (cf. Mason *et al.* 2001; Breeveld *et al.* 2002). These include: the inclusion of a field stop in front of the detector to eliminate stray scattered light from stars outside the detector filed of view; and use of a UV grade MgF coating for the mirrors together with improved contamination control procedures to improve the overall throughput, especially in the UV. Some other differences between the OM and the UVOT are: new ICU circuitry and software for safing the TM (the XMM-Newton mission is a scheduled Guest Observer Facility whereas Swift is a rapidly-pointing, autonomous operations observatory), new DPUs, new DPU software to handle science data, ICU/DPU software to interface with the spacecraft bus, and a new telescope door based on the Triana (Gerstl 1999) door design. A brief comparison between the UVOT, OM, and other UV telescopes is made below.

- Copernicus Princeton Experimental Package (OAO-3 PEP): imaging in the 91.2-327.5 nm range; two photomultiplier tubes for the detector; Cassegrain optics.
- Extreme Ultra-Violet Explorer (EUVE): spectroscopy in the 6.5-70 nm range; three MCPs for the detector; Wolter-Schwarzschild Type II grazing incidence optics.
- Far Ultraviolet Spectroscopic Explorer (FUSE); spectroscopy in the 90.5-119.5 nm range; two double delay-line MCPs for the detector; four off-axis parabolas mirrors.
- Galaxy Evolution Explorer (GALEX): imaging in the 135-280 nm range; two MCPs for the detectors; Ritchey-Chrétien optics.
- International Ultra-Violet Explorer (IUE): echelle spectrographs sensitive in the 115-335 nm range; SEC Vidicon with UV converter for the detector; Ritchey-Chrétien Cassegrain optics.
- Optical Monitor (XMM-Newton OM): imaging and grism spectroscopy in the 170-600 nm range; MCP intensified CCD for the detector; Modified Ritchey-Chrétien optics.
- Ultra-Violet/Optical Telescope (*Swift* UVOT): imaging and grism spectroscopy in the 170-600 nm range; MCP intensified CCD for the detector; Modified Ritchey-Chrétien optics.

## 2.1. Mechanical

The TM is divided into four main sections: the external baffle, the telescope tube, the detector module tube (DMT), and the TM power supply tube. This subdivision allowed interfaces to be minimized and facilitated the alignment of the optics.

The external baffle consists of three tubes each containing internal baffle vanes. At the open end of the external baffle is the UVOT door, which is closed during launch and early orbit to protect the telescope optics. The telescope tube is attached to the external baffle. In this tube are mounted the primary and secondary mirrors. The separation between these two mirrors is tightly controlled, using invar metering rods and heaters, in order to maintain the focus of the telescope. The DMT, behind the telescope tube, contains the external interface to the *Swift* observatory and supports the rest of the optical system on a bulkhead. This bulkhead acts as the optical bench. On the telescope side of the bulkhead are the beam steering mirror (BSM) mechanism, two identical filter wheel mechanisms, two detector assemblies and two high voltage units. On the other side of the bulkhead are two sets of processing electronics. A second bulkhead, mounted between the two remaining tubes, carries the TM Power Supply Unit (TMPSU), which contains the motor drive circuits for the filter wheel and BSM mechanisms, and the external connector panel.

The two DEMs are mounted separately from the TM on a radiator bracket on the observatory optical bench. Both DEMs are box structures machined from solid magnesium and are designed to minimize cosmic radiation.

### 2.1.1. Filter Wheel Mechanisms

The two filter wheel mechanisms are identical to each other and are housed in the DMT. Each filter wheel rotates on a stub axle which is mounted to a base plate. Also attached to the base plate is a stepper motor that drives the filter wheel round. Eleven optical elements are placed at equal angles around the filter wheels. A wheel is driven by the pinion on the 4-phase stepper motor shaft, with a gear ratio of 11 to 1. One revolution of the motor, which requires 200 steps, moves the wheel from one optical element to another and 2200 steps will complete one full rotation of the filter wheel itself.

The filter wheel mechanism is driven in open loop mode by counting steps from a known datum position. Coarse and fine position sensors are provided to relocate the datum position should it be lost, and to verify the wheel position after every rotation to confirm that the center of any optical element has been found. The reflective infrared coarse position sensor is fitted to the wheel and gives a true output when the wheel is within about ±15 degrees of the datum position. The infrared fine position sensor, which is used in transmissive mode, is fitted to the rear end of the motor. An occulting disk with a small aperture, through which the sensor looks, is fitted to the rear extension of the motor shaft. It is aligned such that an element will be correctly positioned when the fine sensor gives a true reading *and* the first phase is energized. It is only at the datum position that both the coarse and fine sensors give a true output.

The filter wheel is rotated at a default pull-in speed of 200 Hz, a cruise speed of 420 Hz, and an acceleration of 2000 Hz/sec. These rates are applied when moving from filter to filter or from datum to filter. However, when seeking the datum, the filter wheel is rotated at a constant 200 Hz until the coarse sensor is detected and then at 10 Hz until the fine sensor is seen. Each filter wheel is rated for the equivalent of 50,000 revolutions over its lifetime. Because of this limitation, filter wheel rotations are monitored throughout the mission in order to monitor the usage.

### 2.1.2. Beam Steering Mirror (BSM) Mechanism

The BSM mechanism is contained in the DMT and houses a cylinder to which a flat mirror is placed at 45 degrees in the path of the incoming beam. The purpose of the mechanism is to steer the incoming light beam onto one of the two redundant detector systems. The BSM mechanism is rotated from one position to the other by pulse counting. The final step drives the rotor to a stop where it will be held by a magnetic detent. The mechanism is rotated 180 degrees between the stops by a 4-step per revolution motor geared at 14.5:1. As there is no harm in overdriving the system against this stop, the motor is always driven the maximum number of steps in the specified direction. The default drive frequency is 2 Hz. There are no sensors in the system and the control mechanism is always open loop.

### 2.1.3. Door Module Latch Mechanism

The UVOT door module is designed to minimize contamination within the telescope. The door is held in position by two hinges and a latch. The door module latch mechanism is a mechanically and electrically redundant system with two reacting cup/cones. The latch mechanisms are powered by paraffin driven actuators. When one of the two actuators is powered the shuttle retracts and the latch-plate is released.

### 2.2. Optics

Figure 3 shows some of the components residing in the telescope tube. The telescope is of a modified Ritchey-Chrétien design. It contains a 30 cm primary and 7.2 cm secondary mirror, which are both made from Zerodur®. The optical train has a primary f-ratio of f/2.0 increasing to f/12.72 after the secondary. The primary mirror is mounted on a strong back for stability and the secondary mirror is mounted onto spider arms. To maintain focus the mirrors are separated by thermally stable INVAR metering rods. Mounted behind the secondary mirror is a reference mirror that aids in determining UVOT's boresight. The boresight (the chief ray of the center field point) is near the center of the detector.

For light rejection the TM has internal and external baffles. The first, second, and third baffle tubes comprise the external baffle, all of which are located forward of the secondary mirrors, and help prevent scattered light from reaching the detectors. The internal baffle lines the inner walls of the telescope tube between the primary and secondary mirrors. Secondary and primary baffles also surround the secondary mirror and the hole at the center of the primary, respectively. Behind the primary mirror is the BSM, which directs light to one of the two detectors.

Before the light enters the detector it passes through a filter housed in a filter wheel. Each filter wheel contains the following elements: blocked position for detector safety, UV-grism, UVW2-filter, V-filter, UVM2-filter, optical-grism, UVW1-filter, U-filter, 4x-magnifier, B-filter, and White-light-filter. The characteristics of the UVOT lenticular color, magnifier, and white-light filters can be found in Table 2. The first column in the table lists the filters. The second and third columns list the central wavelength and full-width-half-max (FWHM) of each filter, respectively. The fourth and fifth columns list the 50% encircled energy radius calculated from data obtained during instrument ground calibration for the 'A' and 'B' channels, respectively. The encircled energy includes the contribution from a collimator; therefore, these values are an upper limit for the UVOT. The responses of the lenticular color and white-light filters are found in Figure 4. The anticipated grism profiles are found in Figure 5.

The grisms provide a low spectral resolution while the magnifier offers a 4x increase in the image scale increasing the f-ratio to f/54 in the blue and providing diffraction-limited images. The glass in the magnifier does not transmit UV light. Because the focal plane is curved, the filters are weakly figured and the surface of the detector window is concave.

| Filter | $\lambda_c$ (nm) | FWHM (nm) | $EE_{A\text{-Channel}}$ (arcsec) | $EE_{B\text{-Channel}}$ (arcsec) |
|---|---|---|---|---|
| V | 544 | 75.0 | 0.62 | 0.59 |
| B | 439 | 98.0 | 0.66 | 0.70 |
| U | 345 | 87.5 | 0.64 | 0.76 |
| UVW1 | 251 | 70.0 | 0.80 | 0.83 |
| UVM2 | 217 | 51.0 | 0.82 | 0.88 |
| UVW2 | 188 | 76.0 | 0.87 | 0.88 |
| White | 385 | 260.0 | 0.53 | 0.56 |

**Table 2. UVOT Lenticular Color, Magnifier, and White-light Filter Characteristics**

## 2.3. Detectors

The two detector assemblies are housed in the DMT. Each detector assembly (Fordham *et al*. 1992; Kawakami *et al*. 1994) consists of detector window that is slightly figured, an S20 photocathode, two stacked Micro-Channel Plates (MCPs), a phosphor screen, tapered fiber-optics, and a CCD (see Figure 6). The photocathode is optimized for the UV and blue wavelengths. The CCD has 385 x 288 pixels, 256 x 256 of which are usable for science observations. Each pixel has a size of 4 x 4 $arcsec^2$ on the sky affording a 17 x 17 $arcmin^2$ field-of-view (FOV). The first MCP pore sizes are 8 μm with distances of 10 μm between pore centers. The second MCP has a pore size of 10 μm with a distance of 12 μm between pore centers.

Photons arriving from the BSM enter the detector window and strike the photocathode. Electrons discharged from the photocathode are driven onto the MCP stack by a bias voltage, and then amplified by the MCPs creating an electron cloud. This electron cloud illuminates the phosphor screen. The photons created from the phosphor screen are directed to the CCD via the fiber-optics, which compensates for the different physical size of the MCPs and CCD. The combination of the MCPs and the CCD provides an amplification of ~$10^6$ of the original signal. The registering of the photon splash caused by an incoming photon is achieved by reading out the CCD at a high frame rate and calculating the centroid of the photon splash's position by means of a real-time algorithm (Michel, Fordham, & Kawakami 1997). The centroiding algorithm yields a large image format for the detector by sub-sampling each of the 256 x 256 CCD pixels into 8 x 8 virtual pixels, thus providing an array of 2048 x 2048 virtual pixels with a size of 0.5 x 0.5 $arcsec^2$ on the sky. A relatively simple centroiding algorithm is used in order to achieve the required processing speed. A consequence of this is that the effective size of the virtual sub-pixels depends (in a repeatable way) on the position within the CCD pixel. An average correction for this effect is applied on board. However, because of small gain changes over the face of the detector, a faint residual pattern remains. This represents a redistribution of photons on an 8 x 8 virtual pixel scale (photon events are conserved) and can be cosmetically removed by ground processing. The frame rate of the UVOT detectors is 10.8 ms for a full 17 x 17 $arcmin^2$ frame.

As with all photon-counting detectors, there is a maximum count rate threshold beyond which coincidence losses become important. For the UVOT this count rate is ~20 counts/s (see Figure 7), assuming a point source; a count rate correction needs to be applied. Details of this count rate correcting can be found elsewhere (Antokhin *et al*, 2002). A CCD dead time correction also needs to be applied during the data processing. Because the local sensitivity of the photocathode and MCPs deteriorates as a function of their total illumination history, care must be taken when observing bright objects. This is accomplished through autonomous onboard code that limits the time spent on these bright sources (see Sections 2.4.2.2.4 and 2.6.1.1). The detector's dark noise is extremely low (approximately $4 \times 10^{-5}$ counts/s/pixel) and can be ignored when compared to other sources of background noise.

## 2.4. Electronics

Figure 8 summarizes the electronic architecture of the UVOT. The ICU controls and monitors the TM via the instrument control bus (ICB). The TM contains the TMPSU that controls (a) a filter wheel, the position of which is monitored by LED illuminated sensors, (b) the heaters and (c) a beam deflector that switches the optical path between the prime and redundant halves of the instrument. The Detector Processing Electronics (DPE) acquires and forwards photon events to the DPU in the DEM. The DPE also (1) controls the high voltage unit attached to the image intensifier, (2) activates calibration flood LEDs, (3) powers to the filter wheel sensors and (4) monitors the effect of the heaters via thermistors. The DPU processes the photon events into lists, images and a parameterized finding chart that are then forwarded directly to the spacecraft.

### 2.4.1. Digital Electronics Module (DEM)

The DEM is the UVOT's high-level command and data processing system. It is comprised of four subsections: the ICU, the DPU, power supply modules, and the Versa Module Eurocard (VME) backplanes (see Figure 9). The DEM power system includes two independent power supply units. Side A of the DEM furnishes +5V, ±12V, and 3.3V to the DPU and side B distributes +5V to the ICU. The dual power supplies have a nominal input voltage of +32V but accommodate between +22V to +35V. Both power supplies provide electromagnetic and radio frequency interference filtering. The DEM chassis, which includes a VME backplane for the DPU and ICU, is based on the SC-9 product.

### 2.4.1.1. Instrument Control Unit (ICU)

The architecture of the ICU is shown in Figure 10. The ICU is responsible for controlling and managing all aspects of the UVOT's operation. These operations include:

- Interacting with the spacecraft to ensure instrument safety during slews
- Autonomous instrument safing in off-nominal observatory conditions
- Emergency communications via the Tracking and Data Relay Satellite System (TDRSS)
- Autonomous protection of the detector from fields containing bright stars using the combination of an on-board star catalogue and signals from a bright source detecting safing circuit
- Interacting with the Figure of Merit computer to select and execute appropriate science observations
- Control and monitoring of instrument thermal state, mechanisms and detector system

ICU capabilities are implemented via a combination of compiled Ada (Booch 1987) code and a customized interpreted scripting language, together with electrically erasable programmable read only memory (EEPROM) located tables of exposure sequences, safety-related information and calibration data.

The programmable read only memory (PROM) contains a bootstrap that, on power-up or reboot, copies the basic set of code from PROM into random access memory (RAM) for execution. The basic code supports basic safing of the instrument, provides housekeeping and thermal control, initiates the loading of operational code into RAM for execution and permits updates to the operational code. EEPROM-A contains the operational code and tables required by the basic code. EEPROM-B contains all remaining tables, the star catalogue and the scripts. RAM consists of 64K words of code space and 64K words of data space.

The ICU watchdog timer gives a timeout after 11 seconds. If the timer reaches zero a power down interrupt is generated and 256 microseconds later, the IC will be reset. This timer is disabled on power up and is enabled by ICU software. The timer can be enabled and disabled by ICU software. This timer is reset after a commandable interval provided the ICU is in communication with the spacecraft. On a less frequent, but also commandable time interval, the counter is reset provided 'aliveness' flags maintained by all software tasks in the ICU are set.

### 2.4.1.2. Data Processing Unit (DPU)

The DPU is a two-card assembly containing the Central Processing Module (CPM) and the *Swift* Communication Module (SCM). The CPM is a RAD 6000™ with 128 MB of RAM and 3 MB of EEPROM. The SCM card is a custom design, which includes a 1553B interface for communication with the *Swift* observatory, a serial Data Capture Interface to receive photon event data from the TM, and SSI circuitry for communication with the ICU. The DPU is a slave to the ICU. Its primary task is to process a moderately high rate of photon event and generate packetized Consultative Committee for Space Data Systems (CCSDS) telemetry products such as images and event lists.

### 2.4.1.3. DEM Interfaces

### 2.4.1.3.1. 1553 Interface

All data (includes commands, housekeeping, science data, time distribution messages, burst alert messages, and inter-instrument communications) is passed to/from the spacecraft and between instruments using a MIL-STD-1553 bus. The spacecraft can accommodate variable length packets in CCSDS packet telemetry format transmitted over the MIL-STD-1553 interface. However, all telemetry packets issued by the UVOT/ICU are of fixed length.

### 2.4.1.3.2. Serial Synchronous Interface (SSI)

The serial synchronous interface (SSI) is a bi-directional communications interface between the DPU and ICU. Both the ICU and the DPU can send and receive data on this interface but the ICU is the master. Commands are sent from the ICU to the DPU. DPU responses are sent to the ICU. They are in data blocks identical in format to spacecraft packets.

The SSI clock frequency is 125 kHz producing a period of 8 us (1 bit-period). The SSI 16-bit data words are separated by at least one bit-period and at most the SSI block gap. The SSI data blocks are separated by at least the SSI block gap (defined in software).

### 2.4.1.3.3. Instrument Control Bus (ICB)

The ICU controls and monitors the telescope module via the instrument control bus (ICB). The ICB is the digital data highway that the ICU uses to send and receive commands and status. An existing standard has been adopted for the ICB called the MACS bus (Modular Attitude Control Systems bus). It is a prioritized multi-master bus.

Because there are a number of units on the bus the ICB has several functions. The detail of the functions performed on the bus is controlled by software in the ICU. The functions performed via the ICB are:

- Loading of tables into the detectors
- Commanding of the detectors
- Status monitoring of detectors
- Reading filter wheel position sensors and temperature sensors
- Controlling power switching
- Controlling heater switching
- Controlling motor drives
- Monitoring voltages/currents

The MACS bus specification defines a redundant bus. Redundancy is provided in the *Swift* UVOT by two separate detector chains, and therefore only one MACS interface is used per redundant half.

### 2.4.1.3.4. Time

Onboard time is managed using two components: spacecraft clock and a universal time correlation factor (UTCF). The spacecraft clock is the spacecraft's internal clock used for the majority of onboard functions (e.g., all CCSDS secondary header time tags, management of stored command processing functions). The spacecraft clock is set at initial power-on and is nominally run without adjustments. The UTCF is a bias that is adjusted such that the sum of the spacecraft clock with the UTCF yields a time that is as close as possible to UTC.

The spacecraft transmits time to all instruments. Spacecraft time and a UTCF are transmitted over the 1553 bus once every second and are valid for the next 1 pulse per second epoch.

### 2.4.2. Telescope Module (TM) Electronics

### 2.4.2.1. Telescope Module Power Supply Unit (TMPSU)

The telescope module power supply (TMPSU) converts the spacecraft power bus to power rails within the telescope module. One set of rails powers the digital and analogue electronics and high voltages. The analogue electronics, in turn, controls the high voltages and powers filter wheel fine sensor LED and flood LED's. The other set power the mechanisms and filter wheel coarse sensor. The integral ICB interface provides the channel for control of the coarse sensor, the flood LED's, the analogue and digital electronics and the return of current, high voltage and fine sensor status values. Additionally the main s/c power, routed via the TMPSU, is used to drive the heaters.

### 2.4.2.2. Detector System

#### 2.4.2.2.1. Camera Head

The sensor in the Camera Head is an English Electric Valve (EEV) CCD-02-06 which is a frame transfer device running with a vertical clock rate of 1.67 MHz and a horizontal readout rate of 10 MHz. The CCD is of well-proven design and is used in many monochrome commercial and scientific TV applications. The dummy output from the CCD is subtracted from the video signal to reduce the level of saturation of the final video amplifier stage. The main cause of this is clock feed-through in the CCD wiring and the reset spike. Figure 11 shows the functional blocks of the camera.

Under control from the DPE, the camera is capable of reading out of a number of windows in the CCD image in rapid succession, or full 256 x 256 pixel frames. The integration time is typically 11 ms.

#### 2.4.2.2.2. High Voltage Control Unit

The High Voltage Control Unit (HVU) comprises three converters. Converters 1 and 2, working in parallel, produce the voltage across the MCP1 bottom plate and the cathode ($V_{cathode}$), and the voltage across MCP1 ($V_{mcp1}$). Converter 1 produces a negative voltage so that with the use of resistive division with converter 2 it obtains a zero volt output for $V_{cathode}$ on command. Converter 3 is in series with converter 1 and 2 and produces the bias voltage across MCP23 and the anode gap voltage known as $V_{mcp23}$ and $V_{anode}$ where $V_{anode}$ is produced by extension of the voltage multiplier chain used to create $V_{mcp23}$.

In order to prevent potential reversal of any intensifier plate the bias voltages must be applied sequentially; this sequence being $V_{anode}/V_{mcp23}$-$V_{mcp1}$ then $V_{cathode}$. The HVU hardware will prevent any controlled static potentials from reverse bias conditions even if commanded to do so.

#### 2.4.2.2.3. Image Intensifier

A discussion of the image intensifier was contained in Section 2.3.

#### 2.4.2.2.4. Detector Safety Circuit

If exposed to bright sources the UVOT detectors may be temporarily or permanently degraded. The main causes are:

- Fluorescence/phosphorescence – this is a temporal effect. If a 5.6 mag A0 star illuminates the detector through the white filter for one minute, then its phosphorescence remains above detection limits (0.008 counts/s) for 16 hours.
- MCP gain loss – this results in permanent performance degradation, due to a localized gain loss in the MCP and a loss in sensitivity of the photocathode due to ion feedback from MCP pores.

The UVOT safety circuit protects the detector against *unexpected* bright sources. The techniques used to limit or prevent observations when bright sources are known to be present are discussed in Section 2.6. This circuit operates independently of the ICU but is under its overall control. It has a fast response time of less than one second to prevent damage to the detector system. The safety circuit is flexible enough to cope with changing circumstances, has the facility to be disabled and is insensitive to the effects of penetrating radiation in the CCD and detector.

During testing of the UVOT image tubes it was observed that in conditions of high event rates when the overall system would, in normal operation, be showing significant coincidence losses, there was a broadening of the event profile. This was found to be predominantly an effect of the image tube. It was considered that this repeatable characteristic provided a good indication of source brightness. Consequently, the safety circuit design was based on measuring the maximum width of a star profile.

The system connects between the camera and the DPE (see Figure 12) and scans the raw video data for a predetermined number of consecutive pixels above a preset amplitude threshold. To ensure immunity to the effects of penetrating radiation these consecutive pixels must be present for multiple consecutive frames. It is also necessary to ensure that only valid pixels from the camera were analyzed. This is achieved by synchronizing the pixel analysis to the frame sync signal issued at the start of each frame.

When the safety circuit detects a bright source, it automatically powers down the cathode voltage of the detector. This significantly reduces detector gain. It also reports the alert signal in the safety circuit status register. The ICU may then take extra safety actions: the MCP bias voltages will be ramped down to 0V, and the filter wheel is moved to the blocked position. For test purposes, the cathode voltage control circuit may be disabled, while still reporting the alert in the control register. A more detailed description of the safety circuit can be found elsewhere (Hancock & Kawakami 2004).

### 2.4.2.2.5. Detector Processing Electronics

The principal features of the detector processing electronics are: generation of the detector head clock sequences to operate the CCD in a frame transfer mode, specification of the areas of the CCD to be read out, event detection, event centroiding, producing engineering data, construction and transmission of data to the DPU, and ICB interface for control of the previous items. A block diagram of the detector processing electronics is shown in Figure 13.

### 2.4.2.2.6. Window Bitmap RAM

Before the detector processing electronics may be used, the window bitmap RAM must be loaded. The RAM is 64k by 4 bits. The information loaded will cause only those CCD pixels within the desired windows to be readout, i.e. a clocking sequence is generated for the desired camera format.

For every location on the CCD, there is a location in RAM. During row readout, the corresponding RAM contents are interpreted as a window ID. An ID in the range of 1 to 15 is a valid window ID and the corresponding pixel pair is readout, whereas a value of zero means that it is not in any window and will not be readout. By loading up the RAM accordingly, the detector area can be thus divided up into a collection of windows of varying size.

#### 2.4.2.2.6.1. Centroid Lookup RAM

Centroiding is the process of locating the position of an event to accuracy greater than that of a CCD pixel. For each event and in the x and y-axes, the processing electronics produces two 8-bit numbers, labeled 'm' and 'n.' The division m/n is the fractional position within a CCD pixel of the event. The range is divided into 8 bins, otherwise known as sub-pixels. Rather than performing this calculation for every event, there are two (64k by 4 bit) tables containing all possible results of the division. The 'm' and 'n' values are combined into a single 16-bit address that is used to lookup the result. The result is in the range 0-7. Preparing the table contents requires two sets of 9 'channel boundary' values giving the edges of the sub-pixels in both x and y. They are in the range -1.00 to 1.00. The tables are loaded from the ICU via the ICB.

#### 2.4.2.2.6.2. Output Data Formats

The output of the processing electronics to the DPU is a series of 24 bit words, one per event processed. The format of the word is determined by the data acquisition mode set via the ICB. There are four scientific modes (numbered 0 to 3) and effectively two engineering modes (numbered 4 to 7). The normal mode for science observation is 3 (High Resolution, Full).

The scientific modes provide event positions in the form of the x and y CCD pixel number, the sub-pixel number in x and y and, for the windowed modes only (0 and 2) the window ID of the window in which they occurred. There are two full frame modes where the window ID is replaced by the most significant bits of the x and y CCD pixel counters, thus giving 16 tiles covering the full detector area.

The engineering modes provide information for setting up and checking the detector. Modes 4 or 5 capture centroiding information in the form of events in which the x and y co-ordinates are replaced with the 'm' and 'n' values. The two 256 by 256 'pseudo images' thus formed can be used to calculate a new sub-pixel channel boundaries from which the centroid lookup table can be reloaded. Note that a) modes 4 and 5 are equivalent and both formats are transmitted at once b) the first X M/N event for each frame is not transmitted. Modes 6 or 7 give event height leading to a 1D image i.e. a histogram. They also produce event energy records in which the energy value is set to zero, due to this feature being removed from the design. Therefore, all records of this format should be ignored. Note that mode 6 and 7 are equivalent and both formats are transmitted at once.

In addition, there are two words of all zeros, the 'frame tags', transmitted at the start of each frame. These are used for frame counting and timing purposes. This feature may be disabled via the ICB.

### 2.5. Thermal

The UVOT's thermal design is based on three instrument requirements: when the UVOT is in its nominal operating mode, the telescope mounting flange is thermally controlled in order to minimize the heat transfer between the TM and the observatory OB; the mirror section of the telescope must be held at $19.5 \pm 0.5$ °C to maintain the mirror separation; and, the temperature of the electronics must be kept within the operating and survival temperature limits prescribed by the manufacturers.

The thermal control for the TM is active. Heaters and thermistors regulate the temperature of the TM interface and the environment surrounding the primary and secondary mirrors. Heat pipes transport the power dissipated from the electronics at the back of the TM to the external baffle at the front of the TM and out to space. For this reason, the external surface of the individual baffle tube closest to the door and the door itself are coated with a high-emissivity, low-absorptivity white paint and the internal surface of the external baffle is coated with high-emissivity black paint. The remaining external surfaces of the TM are covered in multi-layer insulation (MLI) to minimize the heat transfer by radiation between the TM and the space/observatory environment. Figure 14 illustrates the thermal design of the TM.

The DEMS are covered in MLI and bolted to a radiator bracket. There is no active thermal control when the UVOT is on. The power dissipated by the operating DEM keeps the non-operating unit within its temperature limits. The radiator bracket is also coated with the same high-emissivity, low-absorptivity white paint as the TM, to enable heat to be transferred to deep space.

### 2.6. Software

An over view of the UVOT software can be found in Figure 15. Both the instrument control unit (ICU) and the data processing unit (DPU) contain microprocessors running custom flight software.

#### 2.6.1. ICU Software

The software components that reside in the ICU are responsible for the autonomous control of exposures and for maintaining the health and safety of the instrument. All ICU code is written in Ada (Booch 1987) except where speed requirements dictated assembler.

In order to achieve the scientific goals of the UVOT, the ICU is required to autonomously control the instrument during execution of one of two types of exposure sequences: automated and pre-planned, which are described in Sections 3.3 and 3.4 respectively. Further requirements placed on the ICU include: responding to pointing constraints imposed by the Sun, Earth, Moon and planets; avoiding damage to the instrument from known bright sources in, or close to, the FOV; monitoring critical parameters for out-of–limit conditions and taking appropriate recovery action; protecting the instrument during slews or during any loss of spacecraft attitude; recovering from any command failures; responding quickly to an emergency shutdown warning from the spacecraft; and running continuously for at least 72 hours without ground intervention. In order to accommodate changes in observational procedure that could occur over the lifetime of the instrument, the ICU software was designed to be as flexible and as readily re-configurable a system as possible. Further details on the ICU software can be found in Huckle and Smith (2004).

### 2.6.1.1. Accommodating Constraints on Observing

The pointing constraints imposed by the Sun, Moon, and Earth, each of which is very bright and therefore potentially damaging to the detector, generate an avoidance angle to which the satellite must adhere. The *Swift* Observatory will be in a low Earth orbit with an orbital period of approximately 96 minutes. Assuming no additional pointing constraints imposed by the current position of the Sun or Moon, the Earth's avoidance angle implies that it will not be possible to observe any source for longer than about 45 minutes, at which time the satellite must slew to another source. For sources at high elevation above the satellite orbital plane, this maximum observing time is decreased, falling to zero at 84 degrees.

The detector is also susceptible to damage from sources such as stars or planets brighter than about $8^{th}$ magnitude (depending on their color). The ICU must monitor long exposures of any given source and ensure that the total accumulated counts on any given detector location does not exceed a damage limit.

Additionally, due to particle radiation, the instrument needs to be protected during passages through the South Atlantic Anomaly (SAA). These interruptions will occur multiple times in any twenty-four hour period and each may last up to ten minutes. This protection is achieved by ramping down the high voltages controlling the image intensifier.

Due to these pointing constraints, it is impossible to perform a continuous sequence of automated or pre-planned targets. To accommodate these pointing constraints and observational interruptions the ICU design must: allow for automated and planned observations to be interleaved, expect either type of exposure to be interrupted by pointing and SAA constraints, make decisions on exit from interruptions on how to reconfigure the instrument and restart the exposure, and permit the curtailing of all types of exposures to prevent detector damage.

### 2.6.1.2. Intra-Observatory Communications

In order for the ICU to determine its course of action, the following information is supplied from the spacecraft, the Figure of Merit (FOM; the onboard software that determines the relative observing priority of the target), the BAT (Barthelmy 2004), and the XRT (Burrows *et al.* 2004):

- The spacecraft supplies an attitude control system message at a frequency of 5 Hz. This contains an identifier uniquely specifying the current observation. It also supplies Boolean flags indicating whether the spacecraft has settled, or is within 10 arc minutes of its final position, and whether the spacecraft is currently inside the SAA. It also includes a flag that, if true, signals that the spacecraft may be about to remove instrument power. Positional information, specifically the current right ascension and declination of the source and the satellite's latitude and longitude are supplied. Timing information, in the form of the current spacecraft clock setting, is contained within the same record.
- The spacecraft supplies a hardware driven 1 Hz timing pulse, referred to as the 1PPS. Between these pulses, a message that supplies the spacecraft clock and UTC values at the next pulse is sent.
- Telecommands are sent by the spacecraft to notify the instrument when a slew is about to commence or if a signaled slew has been abandoned.
- Prior to slewing to the next source, referred to here as the target, the FOM Processor sends out a FONEXTOBSINFO message detailing the next observation. This includes an identifier uniquely specifying the next observation and a flag identifying it as an automated or planned target. For an automated target the time since the source was detected and whether this is the first visit to that source are also given. The target right ascension, declination, and roll are supplied, together with the anticipated maximum observing time on the source before the next interruption, allowing for anticipated interruptions by the SAA. For all types of observations, it further supplies a value known as the UVOT mode that the ICU uses to select the sequence of exposures to run on the target.
- The BAT supplies a message detailing burst brightness.

- The XRT may supply a refined position for a GRB while the UVOT is performing the finding chart exposure. This message impacts the area of detector that needs to be processed to ensure the inclusion of the source.

#### 2.6.1.3. ICU Software Design Philosophy

In order to achieve the flexibility required, a design based primarily around three types of tables stored in EEPROM was chosen. All these tables have an associated Cyclic Redundancy Check (CRC) value that is used to validate each table as it is loaded into RAM prior to use. It is intended that most, if not all, proposed changes to the behavior of the system can be achieved by modifying one or more of these tables. There are three types of table images stored in EEPROM: Relative Time Sequences (RTSs), Data Tables, and Action Tables.

##### 2.6.1.3.1. Relative Time Sequences (RTSs)

RTSs are sequences of command words derived from text files containing scripts. There are two types of RTSs: those containing the translated versions of the RTS scripts and those that index into them. RTSs provide a separate layer of "software" on top of the Ada code. It deals only with the higher-level aspects of UVOT control. Since the Integrated Test and Operations System (ITOS; see http://itos.gsfc.nasa.gov/opendoc/techbrief.html) is used as the *Swift* control and monitoring system, the syntax of the RTSs is by design similar to ITOS procedures. RTSs therefore act as an on-board UVOT command facility into which changes in ideas and circumstances can more readily be incorporated.

Only one (top-level) RTS may run at any time, although it may call other RTSs as subroutines. Each RTS is assigned a priority, which is also used by any RTSs called. If a commanded RTS has a higher priority than the one currently running, the latter will be shut down and the new RTS run in its place. Priorities are set based firstly on safety and then on science considerations. Figure 16 demonstrates how Ada reads an RTS.

##### 2.6.1.3.2. Data Tables

Data tables contain sets of numbers that may need to change in the course of the mission. These include calibration data for the on-board high voltage ramping and heater control algorithms. Two further tables contain the automated and preplanned target exposure configurations. Count rate and avoidance angle tables are also present.

###### 2.6.1.3.2.1. On-Board Catalogue

The ICU software uses an on-board star catalogue to determine the magnitude and color of stars in, or close to, the target FOV. This information is used in two ways to protect the detector from bright source damage:

- If the star is within an EEPROM tabulated angle of avoidance around the FOV for that magnitude, observation of that FOV is prohibited. This prevents stray light (for example, reflected off the baffle) from entering the optics.
- If the star is in the FOV and does not violate any angle of avoidance criteria, its color is used to index into a table giving the theoretical count rate as a function of filter. The count rate is then scaled by the catalogued magnitude. The count rate for each star thus calculated is used to decide how long, if at all, an observation at a particular pointing, in that filter, can safely continue. The color index is a B-V magnitude.

The catalogue is stored in EEPROM and is divided into three contiguous sections: the main catalogue, the addendum, and the pointer table. The main catalogue contains stars down to $12^{th}$ magnitude. As the detector is more sensitive to blue light than to red, sources that are too red to affect the detector are filtered out from the catalog. Over 200,000 remaining stars are stored, derived from the Tycho II (Hog 2000), GCVS (Kholopov *et al.* 1985a; Kholopov *et al.* 1985b; Kholopov *et al.* 1988; Durlevich *et al.* 1994), NGC (Sulentic & Tifft 1973), and Yale Bright Star (Hoffleit & Jaschek 1991) catalogues. For efficiency of access, the catalogue is split into 2524 sky areas of approximately equal solid angle, in 44 declination bands. Within each area the position of each source is stored to +/- half an arc minute accuracy in the right ascension and declination axes, relative to the origin of the area. Each source's associated magnitude and color information is also stored. Each sky area is followed by a CRC value for memory corruption checking. The addendum is an area left blank in the EEPROM in case any sources were omitted from the catalogue prior to launch. If a source is added to the addendum then it will be put at the beginning of the blank EEPROM area, followed by a marker indicating the end of the main catalogue. The pointer table allows quick access to catalogue data. It holds a set of pointers that, via two levels of direction, point to each sky area. It is CRC protected.

#### 2.6.1.3.2.2. Planetary Positions

In order to protect the instrument from moving celestial objects, the ICU calculates the positions of the Sun, Earth, and Moon as a backup to the protection already provided by the spacecraft. In addition, similar calculations are performed for Venus, Mars, Jupiter, Saturn, Uranus, and Neptune, as the spacecraft does not provide this information. All target FOVs are compared against these positions and an EEPROM located table of angles of avoidance. Any violation of those angles prevents the exposure except in the cases of the fainter planets Uranus and Neptune, which are considered as stars and a maximum observing time deduced instead. The positions are calculated using formula, algorithms, and data given elsewhere (US Naval Observatory 2001; Seidelmann 1992; Meeus 1985; Duffet-Smith 1979).

### 2.6.1.3.3. Action Tables

Action tables define actions that are to take place when an event or combination of on-board values occurs. Those actions are defined by declaring an RTS to be run. They consist of: a state change table, a limit-checking table that states which RTS is to be run when a limit failure for a particular engineering item occurs, and an errors action table that states which RTS is to be run to perform an error recovery action. All telemetry from the ICU and DPU is internally monitored for error messages. When one is detected, the table is then consulted and, if required, the appropriate RTS is run. The design allows for up to 256 such messages.

#### 2.6.1.3.3.1. State Transitions

In order to achieve its goals, the ICU must successfully transition the UVOT between several instrument configuration states. The possible transitions are shown in Figure 17. Each state transition is performed by a RTS. A look-up table located in EEPROM contains a list of those RTSs against the requested transition. For those transitions that may be autonomous it also contains the required state of internal ICU flags for that particular transition to take place. This table is used in two ways: on command and autonomously.

On reception of a telecommand requesting a particular transition that was issued from the ground (on command) or, alternatively, from a running RTS, the table is scanned for a match against both the current state and the requested state. If a match is found, the relevant RTS is selected and executed. If no match is found, the command is rejected.

A number of events will also cause the ICU to select the required state transition autonomously by comparing the table of internal ICU flags against the current values. If a match is found, the RTS is selected and executed. If no match is found, the request is ignored. The events that trigger such a response are: the spacecraft settled flag becoming false, the reception of the FONEXTOBSINFO message, the spacecraft 'within 10 arc minutes of target' flag being set to true, entry or exit from the SAA, and the Safe Hold flag being set to true.

The purpose and activities of each state are summarized below.

- *Basic*: The ICU enters this state on turn-on. The code executed in this state is resident in ROM and therefore cannot be updated. It ensures that the filter wheel is in the block position and that the high voltages are at zero. It maintains housekeeping, on-board autonomous limit checking, thermal control, and enables the watchdog. These latter processes continue to run in all other states. On command to go to the Safe state, it first loads the operational code from EEPROM into RAM and then performs the transition. All subsequent states are supported only by the operational code.
- *Safe*: As its name implies, in this state the instrument is configured to be least susceptible to damage. The filter wheel is in a blocked position and the high voltages (HVs) are set to zero.

- *Idle*: The UVOT is configured to be ready to observe but is awaiting the next slew. As a safety precaution the cathode voltage is held down at zero and the filter wheel may be placed in the blocked position until observations commence.
- *Slewing*: In this state the ICU cleanly shuts down any current observation, prepares the instrument for slewing, informs the spacecraft when ready to slew, selects the next exposure, and performs various safety calculations.
- *Settling*: After a new GRB has entered the UVOT's field of view but before the spacecraft has settled, it is observed by collecting an event list. As the target will be moving rapidly, it is not possible to collect an image.
- *Finding Chart*: If the target is a new GRB, then once the spacecraft has settled, a 100 second exposure is made in a standard filter to produce a finding chart that will be sent to the ground. In many cases, by the end of the exposure, the XRT will have reported an improved position for the source. The ICU uses that information, and the GRB brightness information supplied by the BAT, when it configures for the subsequent Automated Target exposures.
- *Automated Target* (*AT*): During the period while a GRB has no pointing constraints the ICU will configure and run a series of exposures in this state.
- *Planned Target* (*PT*): The ICU will configure this type of exposure when the FOM informs the ICU, via the FONEXTOBSINFO record, that the spacecraft is performing a planned observation.
- *Safe Pointing* (*SP*): The ICU will configure this when informed that the spacecraft is slewing to a safe pointing. This type of exposure occurs when no automated or planned targets are available.
- *South Atlantic Anomaly (SAA)*: In this configuration, the instrument is configured to be safe whilst passing through the SAA. In particular, the MCP bias voltage is held at 70% of its nominal value and the cathode voltage is at zero.

### 2.6.1.3.3.2. Limit Checking

An on-board algorithm monitors various safety critical items at all times. The information in this table controls the algorithm and consists of the: items to be monitored, frequency of monitoring and phasing of that monitoring within the cycle, valid range of data in raw un-calibrated units, number of times the range must be exceeded before it is considered an error, and recovery action to perform in the event of an error.

Limit checking is done to monitor the voltages, current, temperatures, and the safety circuit. Since testing every possible combination of the software is not possible, hardware watchdog circuits, software task watchdogs, Ada exception handlers, and error actions have been included in order to protect the instrument.

### 2.6.1.4. Overall Data Flow

Figure 18 illustrates how data flows between the software modules that make up the autonomous system of UVOT. The components of the autonomous system are the command distributor, the observation manager, the bright objects manager, limit checking, the telemetry queue manager, and the RTS manager. A description of each is found below.

- *Command Distributor*: This module receives not only all the commands and messages sent on the spacecraft bus, but also all commands internally generated by the RTS Manager. It then distributes them to the appropriate software module.
- *Observation Manager*: This module monitors the spacecraft and FOM messages and maintains a record of the status of the ICU. It uses this information to determine when it is appropriate to issue a suitable RTS command. It accesses the state management tables as well as the AT and PT configuration tables described above as part of this process.
- *Bright Objects Manager*: This returns information about planets and stars near or in the target field of view, using the star catalogue and avoidance angle tables described above.
- *Limit Checking*: This monitors critical engineering values. Using the table described above, it issues requests, if necessary, for a RTS to perform recovery actions from a limit failure.

- *Telemetry Queue Manager*: This monitors all outgoing telemetry for error messages and, using the table described above, issues a request to run a recovery RTS if required.
- *RTS Manager*: This module executes the RTS. It consists of the virtual CPU code to execute the RTS, together with software to scan the RTS index image to permit rapid location of a given RTS.

### 2.6.2. DPU Software

The DPU software is built upon a VxWorks™ Real-Time Operating System. Bootstrap and device driver software for the SSI, Data Capture Interface, 1553 interfaces were developed by SwRI. Application-level science "Data Processing Algorithms" software was developed by Penn State University (PSU). This layered approach to software development was critical to completing the software on time and under budget, as it provided a clear division of labor between the two institutions (see Figure 19). The primary functions of the DPU Flight Software are:

- Receive and execute commands from the ICU and the spacecraft
- Receive and process detector events from the UVOT Telescope Module
- Perform loss-less data compression of image and/or photon event data
- On-board source detection for generation of UV/Optical finding chart
- Generate science data telemetry products
- Generate engineering data telemetry products
- Forward packetized compressed data to the spacecraft Solid State Recorder
- Maintain time synchronization with the spacecraft.

The DPU communicates with the ICU through the SSI, and receives raw photon position and timing data from detector electronics across a serial Data Capture Interface. Because the amount of photon event data that can be collected exceeds the UVOT telemetry allocation, the DPU employs data compression to reduce the size of its telemetry data products. The DPU formats data as CCSDS Source Packets, and forwards telemetry to the Spacecraft Control Unit (SCU) through a MIL-STD-1553 (1553) interface. Housekeeping and science telemetry timestamps are synchronized with the spacecraft clock.

## 3. OBSERVING SCENARIOS

There are five observing scenarios for the UVOT: settling, finding chart, automated targets, pre-planned targets, and safe pointing targets. In order to protect the UVOT detector from damage due to bright sources and charged particles, high voltages are turned down and no observations are performed during slews or SAA transients.

### 3.1. Settling

Once the spacecraft provides notification that a new GRB is within ten arcminutes of the target position, the UVOT commences observing. Photons are collected in event mode over the entire UVOT FOV with, by default, the UVW2 filter. The collection of photons continues in this phase until the spacecraft is considered settled. During this settling period pointing errors are off-nominal, i.e., the target is moving rapidly across the FOV as the spacecraft settles. The data collected is sent to the ground through the Malindi ground station.

### 3.2. Finding Chart

If the intended target is a new GRB, and once the spacecraft is settled, the UVOT begins a 100-second exposure in the V filter to produce a finding chart and a GRB neighborhood image (GeNI). Photons are collected in event mode over the entire UVOT FOV during this phase. Upon completion and processing of the 100-second exposure, a parameterized version of the finding chart and the complete GeNI are sent to the GCN via TDRSS to aid ground-based astronomers in follow-up observations of GRBs. The complete finding chart is later sent through the Malindi ground station. The positional accuracy of the finding chart will be approximately 0.3 arcsec relative to the background stars in the FOV. It is expected that for most bursts the XRT will have reported a better than 5 arcsec position for the target before the end of the finding chart observation. The BAT's positional accuracy will be one to four arcminutes. Figure 20 illustrates the positional accuracy and timing of each *Swift* telescope.

### 3.3. Automated Targets

Once the finding chart and GeNI have been completed, an automated series of exposures using an ordered sequence of filters is executed. The photons are collected in event and image modes with FOVs ranging from eight to twelve arcminutes on a side. Exposure times vary from 10-1000 seconds. The series is based on the optical decay profile of the GRB afterglow and time since the initial burst trigger. Currently, two automated sequences will be launched: bright and dim GRB series. The bright series includes the UV and optical grisms as well as the broadband filters while the dim sequence only contains the broadband filters. Figure 21 demonstrates the bright sequence in which the GRB is observed almost immediately. Figure 22 exhibits a dim sequence in which observations of the GRB are delayed due to some constraint. BB, UU, W1, GV, M2, VV, W2, & GU in the figures are the B, U, UVW1, Visual Grism, UVM2, V, UVW2, & UV Grism filters respectively. The gaps in the figures are due to earth occultation. Only these two series will be loaded at launch. However, new series can be added and existing ones modified as GRB afterglows are better understood.

### 3.4. Pre-Planned Targets

When there is no automated target to observe, observations of planned targets are performed. These pre-planned targets, which include previous automated targets, targets-of-opportunity, and survey targets, are uploaded to the spacecraft from the Mission Operations Center (MOC). Photons can be collected in event and/or image modes with FOVs ranging from one to seventeen arcminutes on a side and with any of the filters in front of the detector window. Exposure times vary from 10-1000 seconds. Collected data are sent to the ground through the Malindi ground station.

### 3.5. Safe Pointing Targets

When observing constraints do not allow observations of automated or pre-planned targets the spacecraft points to predetermined locations on the sky that are observationally safe for the UVOT. Observations can be performed at these locations but are currently part of the baseline.

## 4. INSTRUMENT PRODUCED DATA PRODUCTS

The UVOT generates seven categories of data products: event lists, images, parameterized finding charts, GeNIs, intensifier characteristics, channel boundaries, and centroid confirmation images. Of the seven data products, parameterized finding charts and GeNIs are sent through TDRSS to the GCN; the remaining data products are sent to the ground through the Malindi ground station.

### 4.1. Malindi Data Products

#### 4.1.1. Event Lists

Each photon is reported by its position on the detector and the associated CCD frame time-stamp. The accuracy of photon arrival time is limited by the CCD frame integration period (11ms in full-frame mode). A phase diagram of the cataclysmic variable WW Hor, created from data taken with the XMM-OM, other XMM instruments, and South African Astronomical Observatory (SAAO), demonstrates the type of data that can be expected from the UVOT on orbit (see Figure 23).

#### 4.1.2. Images

An image is a 2-D histogram of the event stream integrated over the specified exposure time (typically 10-1000s). Image pixels can be binned 1x1, 2x2, or 4x4 detector sub-pixels. For a large number of events, image data demands considerably less telemetry than event data but at the expense of greatly reduced time resolution. Figure 24 is an optical grism image that demonstrates the grism data anticipated from the UVOT on orbit. Figure 25 is a composite

UV image taken in the UVM2, UVW1, and U filters that demonstrates the images expected from the UVOT on orbit. Both images were taken with the XMM-OM.

### 4.1.3. Intensifier Characteristics

Intensifier Characteristic data are a histogram of the "pulse height" of a set of events acquired with an internal calibration lamp. Examination of these data aids in adjusting the HV detector gains and in checking for any effects of aging.

### 4.1.4. Channel Boundaries

Channel Boundaries data are a set of calibration threshold values which, when loaded into the detector electronics, generate the optimal sub-pixel positioning calculation. The detector data utilized in the computation of Channel Boundaries are obtained with an internal calibration lamp.

### 4.1.5. Centroid Confirmation Image

Once the Channel Boundaries have been determined, a Centroid Confirmation Image can be produced. A Centroid Confirmation Image is a 2-D histogram that characterizes the bias of the sub-pixel positioning calculation across 64 regions of the detector. Figure 26 is an actual Centroid Confirmation Image from the UVOT.

## 4.2. TDRSS Data Products

### 4.2.1. Parameterized Finding Chart

The UVOT Finding Chart is not a classical pixelized image; due to the mission requirement that these data product be made available quickly, a list of interesting image pixels and their neighboring pixels is transmitted via TDRSS and converted in ground processing into a parameterized finding chart, a list of the brightest sources in the field and their basic properties. Notably, the GRB position may not be present in this data product if it is too faint relative to field stars! The information sent by the UVOT flight software to the ground via TDRSS is a complicated abbreviation of the full finding chart image and is best understood by examining the steps used to construct it.

- A real-valued global background level for the Finding Chart image is estimated by computing a robust mean from 1000 pixels spread across the field.
- An integer-valued Detection Threshold is chosen such that the Poisson probability of a background pixel exceeding the threshold is $\leq 1 \times 10^{-6}$.
- The Finding Chart image is searched to identify isolated local maxima that exceed the Detection Threshold. The algorithm is not perfect -- in particular for sources with large extent above the threshold (e.g. very bright point sources, or truly extended sources) the algorithm will falsely claim there are multiple local maxima. In this discussion we'll refer to local maxima positions produced by the algorithm as "Finding Chart entries" to emphasize that they do not always correspond one-to-one to stars.
- The number of total counts associated with each entry is estimated.
- Image pixels surrounding the Finding Chart entries are telemetered via TDRSS. The UVOT was given a fixed allocation of 2000 bytes for the Finding Chart message. In order to maximize the scientific value of that limited telemetry, information about the entries is sent in the following order:

    o The entries are sorted by their brightness.
    o HIGH priority data, namely the position and central pixel, are sent for a maximum of 190 entries.
    o MEDIUM priority data, namely the pixels just north, east, south, and west of the central pixel, are sent for as many entries as will fit in the telemetry allocation.
    o LOW priority data, namely the remaining sixteen closest pixels around the center pixel, are sent for as many entries as will fit in the telemetry allocation.

- Thus, for fields containing <41 finding chart entries the ground will get a 5 x 5 (minus corners) postage stamp around each entry. Position and photometry can be estimated by the ground system for all field stars and the GRB, if detected, using 21 pixel neighborhoods. For crowded fields (~>137 entries) the ground will get the central pixels for 190 entries, and the four neighboring pixels for at least 98 of the brightest entries. Position and photometry can be estimated for at least 98 field stars, and for the GRB if ranked sufficiently high in brightness, using 5-pixel neighborhoods. Coarse position and photometry using only the central pixel are available for the dimmest field stars and for the GRB if ranked low in brightness.
- The ground system will insert all the pixels included in the Finding Chart telemetry into a blank Finding Chart image. The result is a "sparse" image, with only the "interesting" (bright) pixels represented.
- The ground system will detect point sources in this image, estimating position and brightness with whatever pixels are available around the source. Care must be taken to appropriately account for missing data in the position and photometry estimates.
- The ground system will adjust the detector coordinates of the point sources to account for the UVOT detector's field distortion.
- The ground system will match the catalog of point sources to an astrometric catalog in order to assign astrometry to the Finding Chart image.
- The ground system will attempt to identify the likely GRB candidate.

Figure 27 is an example of the UVOT finding chart generation. The image on the left was an image taken by the UVOT of a grid mask during instrument calibration. The green boxes mark the 5x5 postage stamps that are telemetered to the ground. The image on the right is the reconstructed parameterized finding chart.

### 4.2.2. GRB Neighborhood Image (GeNI)

The UVOT finding chart only sends down information about the brightest sources in the field. In crowded fields, or for observations in which the GRB afterglow is faint in the V filter, it is possible that no information about the afterglow will arrive at the ground. This would be a frustrating situation for ground-based observers eager to get some follow-up data on the afterglow. While the XRT error circle will be helpful at a minimum level, the UVOT instrument produces a second TDRSS message that will allow ground observers to obtain a wealth of data about the afterglow candidate using a small amount of extra telemetry. The second TDRSS packet contains a 40 x 40 arcsecond unbinned image of the 100-sec V-band exposure used to create the Finding Chart, centered on the XRT position. If no XRT position is available, a 320 x 320 arcsecond image, binned 8 x 8, centered on the BAT position is generated.

The first finding chart message will be processed and distributed via the GCN without waiting for the GeNI data, to meet the requirement of a timely finding chart. This will also result in an accurate astrometry solution for the GeNI. Once the GeNI is received, it will be converted to detector, and then sky coordinates, using the position and binning information in the packet itself, and the refined astrometry solution from the star identification step in the overall finding chart. Using the instrument point spread function (PSF), the position of any sources in the GeNI will be measured. These positions will be compared to the catalog of known sources to determine if a source in the GeNI corresponds to a known source. If an unknown source is found in this box then its position will be calculated. From the exposure time and count rate, its magnitude will be calculated. Products from this analysis can then be: afterglow identification, afterglow position, afterglow magnitude. A refined finding chart image will also be produced, with the GeNI placed in the appropriate location, again with the XRT error circle in place, for distribution. A sample GeNI is provided in Figure 28.

## 5. CALIBRATION

The pre-launch, ground calibration of the UVOT was performed at the Diffraction Grating Evaluation Facility (DGEF) located at NASA's Goddard Space Flight Center (GSFC) and was completed in November, 2002. The following activities were carried out as part of the calibration program: assessing the linearity and dynamic range of the detector's response, quantifying the system's PSF, calculating the system's effective area, determining the detector's flat field response, measuring the distortion in the detector, and characterizing the grisms. A discussion of the specialized optical ground support equipment (GSE) employed during the calibration program and the ground calibration results are found elsewhere (Mason *et al*. 2004; Quijada *et al*. 2004). Results from the on-orbit calibrations program will be published at a later time.

In order that the detector may be calibrated in flight, four flood-LEDs are provided. They are located off-axis, close to the detector, and are positioned so that their focused emission falls on the backside of the blocked filter - facing the detector. The blocked filter is used because it acts as a defocused 'screen' providing the flat field. The LEDs primarily emit in the green but they also emit in the UV range. Their intensity is controlled via ICB commands routed from the Detector analogue control card to a 4-bit port, thus allowing 16 possible levels. The LEDs are driven in such a way that if one fails the remaining LEDs will remain fully functional.

## 6. UVOT SCIENCE DATA PRODUCTS

UVOT telemetry is received from the MOC by the *Swift* Data Center (SDC) and converted into FITS files. UVOT science data are organized according to five criteria: data mode, filter, on-chip binning, window size, and window location. Level-1 FITS files contain raw images and unscreened event lists.

Level-2 FITS files contain calibrated sky images and event lists that have undergone a number of data reduction steps and standard screening procedures in the SDC pipeline. It is envisaged that the majority of science users will begin their data analysis using the Level-2 products. It would be advantageous to begin at Level-I if there have been some updates to the calibration files in the caldb since the last time the data were reprocessed, or if the science would benefit from non-standard screening.

Level-3 data products contain higher-level science files consisting of detected source lists cross-calibrated with optical source catalogues, composite images, and flux-calibrated source light curves. Specialized tools are required to produce grism products: *uvotimgrism* extracts spectra from images and *uvotevgrism* adds a wavelength column to the event lists. Level-3 products will be produced by the pipeline and archived.

### 6.1. Image Data

A suite of software tools have been developed in order to obtain Level-2 images from Level-1 FITS files: *uvotbadpix*, *uvotmodmap*, *uvotflatfield*, *swiftrans*, and *uvotexpmap*.

- The *uvotbadpix* tool creates a quality map that flags bad pixels using data from two sources. The first is a list of dead, hot, and flickering pixels stored in the *Swift* caldb area. The second is the individual images themselves, where an inspection of the pixel values reveals whether any pixels have been corrupted by the UVOT on-board image compression algorithm.
- Raw images contain a modulo 8 fixed-pattern which is a by-product of the on-board centroiding algorithm. The pattern is removed by correcting sub-pixel sizes to a linear array using *uvotmodmap*.
- Images are corrected for pixel sensitivity variations by dividing through by a flat-field image using *uvotflatfield*. The flat-field image is stored in the caldb.
- The *swiftrans* tool converts raw pixel coordinates to either detector or sky coordinates by applying corrections for instrument distortion and boresight. Distortion and boresight correction factors are stored in the caldb. Although the majority of Level-2 images are stored in sky coordinates, grism images are also provided in detector coordinates, which is the appropriate system in which to extract grism spectra.
- Exposure maps for each image are created with *uvotexpmap* using attitude and distortion data and the quality maps created with the *uvotbadpix* tool.

### 6.2. Event Data

Level-1 UVOT event files undergo the following reduction steps to be converted to Level-2 files: *coordinator* and *uvotscreen*.

- The *coordinator* task populates the detector (DETX, DETY) and sky (X and Y) pixel coordinate columns. It corrects for detector and optics distortion, boresight offset, and spacecraft drift.

- Based on time-tagged, orbit, and attitude data contained in the Level-I housekeeping files, *uvotscreen* determines time intervals of bad data, removing the appropriate rows from the events table. It also rejects events damaged by data compression or resident within a bad pixel.

# 7. UVOT SCIENCE GOALS

The top level science goals of the UVOT are to capture the early light of GRBs, rapidly determine their positions to sub-arcsecond accuracy, quickly follow-up GRB afterglows, identify the GRB environment, provide spectral or photometric redshifts, and provide timing analysis of GRB afterglows. The UVOT will also be key in unraveling the question of dark bursts.

## 7.1. Early Light of GRBs

Because of the rapid slewing capabilities of *Swift*, the UVOT will be able to capture the early (~30s) UV photons of GRBs during the settling phase (see Section 3.1), similar to ROTSE's early observations of GRB 990123 (Akerlof *et al.* 1999). All UV photons in the entire 17 x 17 arcmin$^2$ FOV will be recorded in an event list.

## 7.2. Rapid GRB Positions

Due to the rapid decline of the afterglow's light-curve, localizing the GRB's position is critical for follow-up observations. The UVOT provides a finding chart as described in Section 3.2 and 4.2.1 above. This finding chart is telemetered to the ground in order to assist ground-based observers in pinpointing the burst. This is typically accomplished in less than 300 seconds while the afterglow is still bright.

## 7.3. Rapid Afterglow Follow-up & Identifying GRB Environment

After the finding chart exposure has ended, the UVOT begins an automated observation sequence (see Section 3.3). For bright bursts the filter sequence includes grisms. For fainter targets (17.0 < $m_V$ < 24.0), light curves are acquired by cycling through the six broadband filters. Source variability during exposures can be monitored by collecting data in event mode.

*Swift* follow-up of the afterglow is important to our understanding of the GRB environment. Rapid follow-up of GRBs 990123 and 021004 revealed that the afterglow emission was due to reverse external (possibly internal; Mészáros & Rees 1999) and superimposed forward-reverse shocks (Kobayashi & Zhang 2003), respectively. The UVOT will be able to quickly follow-up many afterglows, which will enlarge the current sample significantly. A large database will allow the location and mechanism of the prompt emission to be identified (Roming *et al.* 2001; Hunsberger *et al.* 2003). It will also constrain the Lorentz factor, which is a crucial GRB modeling parameter (Zhang, Kobayashi, & Mészáros 2003). The image magnifier can be used to determine an accurate position of the GRB within its host galaxy and to detect GRB afterglows to extremely faint magnitudes (the magnifier transmits photons in the 300-650 nm band).

Rapid UVOT follow-up will also facilitate solving the question of "dark" GRBs. Ninety percent of GRBs have a detectable x-ray afterglow. In contrast, sixty percent of long GRBs, localized with *BeppoSAX*, lack a detectable optical afterglow (Lazzati, Covino, & Ghisellini 2002). However, of the fourteen *HETE-2* bursts localized with the Wide-field X-ray Monitor (WXM) and Soft X-ray Camera (SXC), thirteen have had optical transients (Lamb 2003) indicating that there are dark bursts which are classified as such only because of poor localizations and slow follow-up. Currently there are three hypotheses to explain dark bursts (Lamb 2003): there is dust extinction in the vicinity of the GRB (cf. Reichart & Price 2003; Klose *et al.* 2003; Guidorzi *et al.* 2003; Piro *et al.* 2002), the GRB is at a high redshift (cf. Lamb & Reichart 2000; Weinberg *et al.* 2003), or the GRB is intrinsically faint (cf. Lazzati, Covino, & Ghisellini, 2002; Berger *et al.* 2002; Lamb *et al.* 2003; Pandey *et al.* 2003). In conjunction with ground-based telescopes, the UVOT will address the following questions about dark bursts:

- Are all long bursts intrinsically the same but due to dust extinction or high redshifts some are manifested as optically dark/dim?

- Are 'dark bursts' optically faint at all times, or do they simply fade more rapidly than other bursts?
- Are some bursts fundamentally different?
- What percentage of bursts is truly dark?
- Can dark bursts be used to determine how many times reionization has occurred in the Universe?

### 7.4. Spectral and Photometric Redshifts

GRBs are expected to occur out to at least $z \approx 10$ and possibly as high as $z \approx 15\text{-}20$ (Lamb & Reichart 2000). Since the GRB redshift distribution is approximately proportional to the star formation rate (Weinberg *et al.* 2003), it is expected that 10-40% of GRBs should occur beyond $z = 5$. At these redshifts the Lyman limit lies longward of the optical; therefore, these GRBs will be optically dark to UVOT. However, for redshifts in the $1.5 < z < 4.5$ range, UVOT should be able to detect the Lyman limit. For bright bursts the UVOT can measure the redshift using the optical and UV grisms. For fainter targets ($17.0 < m_B < 24.0$) photometric redshifts can be obtained by comparing fluxes in the broadband filters.

### 7.5. Timing Analysis

The UVOT will begin observing bursts from approximately 35-70 seconds until days after the burst is detected. Light curves of these bursts and their afterglows will be compiled in the different filters.

## 8. CONCLUSIONS

The *Swift* mission was launched on November 20, 2004. The UVOT is an important component to the *Swift* Observatory. Because of *Swift's* ability to rapidly slew to a target, the UVOT should be able to capture the early light from some GRBs and will be a dedicated instrument in the follow-up observations of most afterglows. It is anticipated that the UVOT will be key in unraveling such questions as: what are the GRB fireball properties, what are the progenitors of GRBs, are there other class/sub-classes of GRBs, are "dark" bursts due to obscuring dust or are their redshifts too high for optical observations?

## ACKNOWLEDGMENTS

We are very grateful to the large team of dedicated individuals from PSU, Mullard Space Science Laboratory (MSSL), SwRI, Swales Aerospace, and GSFC who, along with the authors, had a significant role in the designing, building, and testing of the UVOT. It is not feasible to list all of these individuals, however we wish to specifically recognize, in alphabetical order, Pete Altimore, Dave Baran, Ray Boucarut, Louisa Bradley, Courtney Chadwick, Bill Chang, Marg Chester, Paul Connors, Diane Day, Sharissa Feasler, Lynn Gilbert, Mark Hailey, Dan Hein, Jeff Hocker, Ritva Keski-Kuha, Brian Kittle, Shane Lanzendorfer, Tim Madison, Al Mariano, Ian Phillips, Manuel Quijada, Alex Rousseau, Peter Sheather, Jared Shoemaker, Frank Thurlow, Graham Willis, and Tim Zukowski. We would like to extend a special vote of thanks to Renan Borelli for his assistance as the COTR for UVOT. We would also like to thank Mark Cropper and Leisa Townsley for their leadership during the Phase A portion of the *Swift* Project. This work is sponsored at PSU by NASA's Office of Space Science through grant NAG5-8401, and at MSSL by funding from PPARC.

Figure 1. UVOT Placement on the *Swift* Spacecraft (**Color**)

Figure 2. UVOT Schematic (**Color**)

Figure 3. UVOT Telescope Tube Components

Figure 4. UVOT Lenticular Color and White-light Filter Response (**Color**)

Figure 5. UVOT Anticipated Grism Response. Insufficient quality data was obtained during ground calibration to characterize the grism response pre-launch.

Figure 6. UVOT Detector Assembly

Figure 7. UVOT Channel A CCD Linearity (**Color**)

Figure 8. UVOT Electronic Architecture (Only the prime channel is shown for clarity)

Figure 9. Digital Electronics Module (DEM) Electronic Architecture

Figure 10. Instrument Control Unit (ICU) Electronic Architecture

Figure 11. Block Diagram of Camera Head Electronics

Figure 12. Safety Circuit Connections to Detector Processing Electronics (DPE) and Camera

Figure 13. Detector Processing Electronics

Figure 14. UVOT Telescope Module (TM) Thermal Design

Figure 15. UVOT Software Overview

Figure 16. The Virtual CPU Flow Diagram (For clarity, not all possible command codes are shown)

Figure 17. UVOT Possible State Transitions

Figure 18. Overview of Autonomous System

Figure 19. Data Processing Unit (DPU) Software Architecture

Figure 20. Rapid position determination of GRBs is made using the three *Swift* telescopes. Alerts and images from each instrument are required to be delivered in the times listed. (**Color**)

Figure 21. UVOT Bright GRB Observation Sequence - Immediate GRB Observation (**Color**)

Figure 22. UVOT Dim GRB Observation Sequence - Delayed GRB Observation (**Color**)

Figure 23. CV WW Hor taken with the XMM-OM, other XMM instruments, and SAAO typifying the type of timing data expected for UVOT

Figure 24. Optical grism image taken with the XMM-OM. This image is indicative of the grism data anticipated from the UVOT on orbit

Figure 25. A composite UV image taken in the UVM2, UVW1, and U filters with the XMM-OM. This image is indicative of the images expected from the UVOT on orbit (**Color**)

Figure 26. Centroid Confirmation Image from the UVOT

Figure 27. UVOT Finding Chart generated by the UVOT. The image on the left was an image taken by the UVOT of a grid mask during instrument calibration. The green boxes mark the 5x5 postage stamps that are telemetered to the ground. The image on the right is the reconstructed parameterized finding chart.

Figure 28. Sample UVOT GRB Neighborhood Image (GeNI). The image on the left is the finding chart image taken by the UVOT of a star mask during instrument calibration. The image on the right is a sample GeNI.

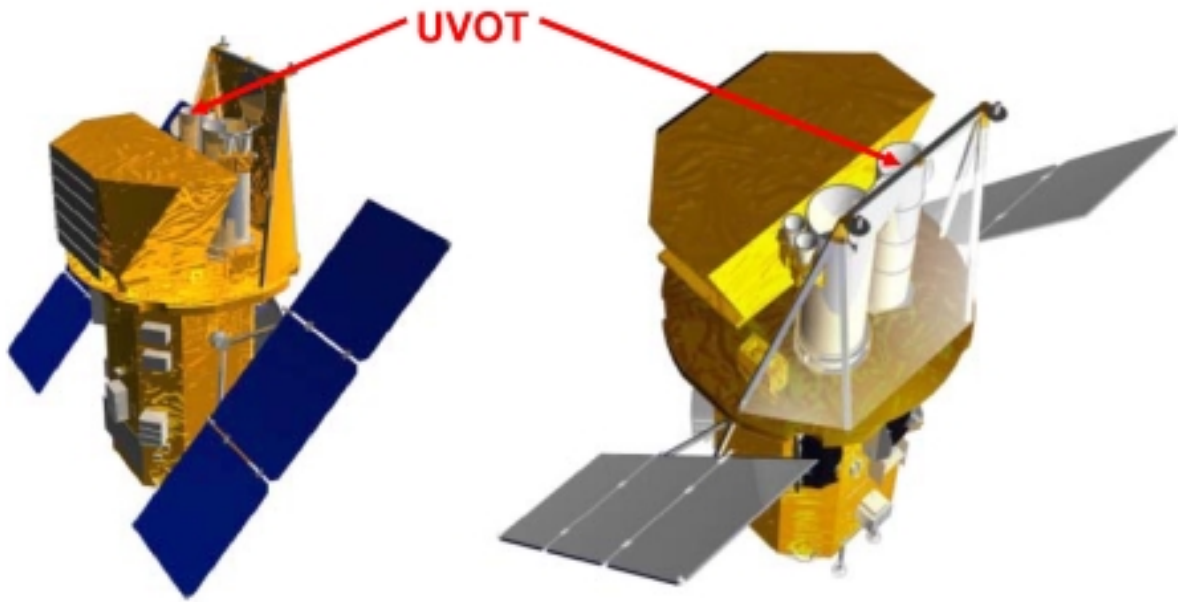

Figure 1

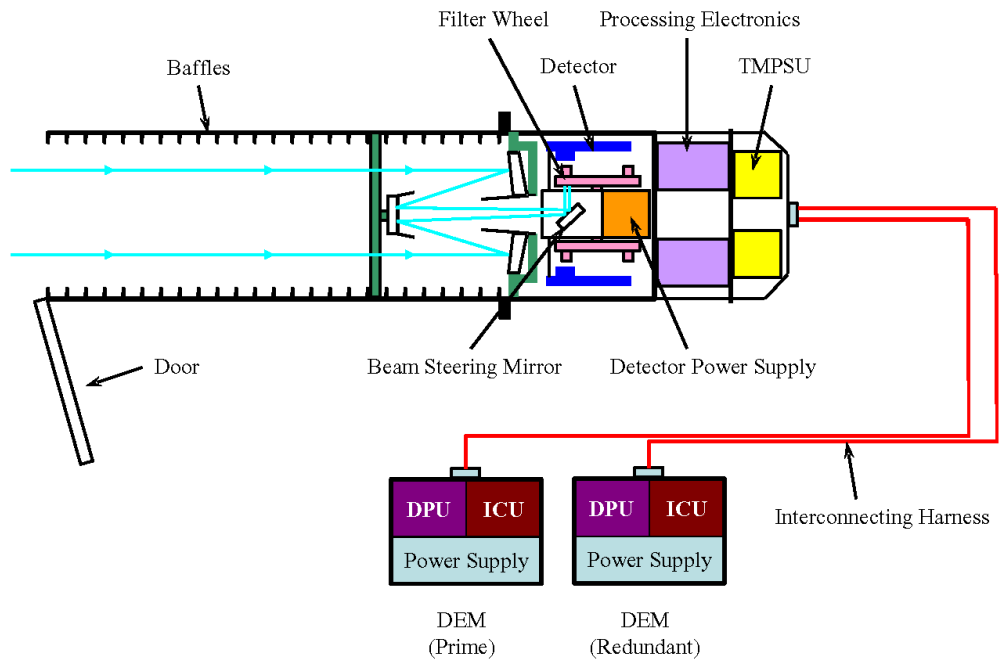

Figure 2

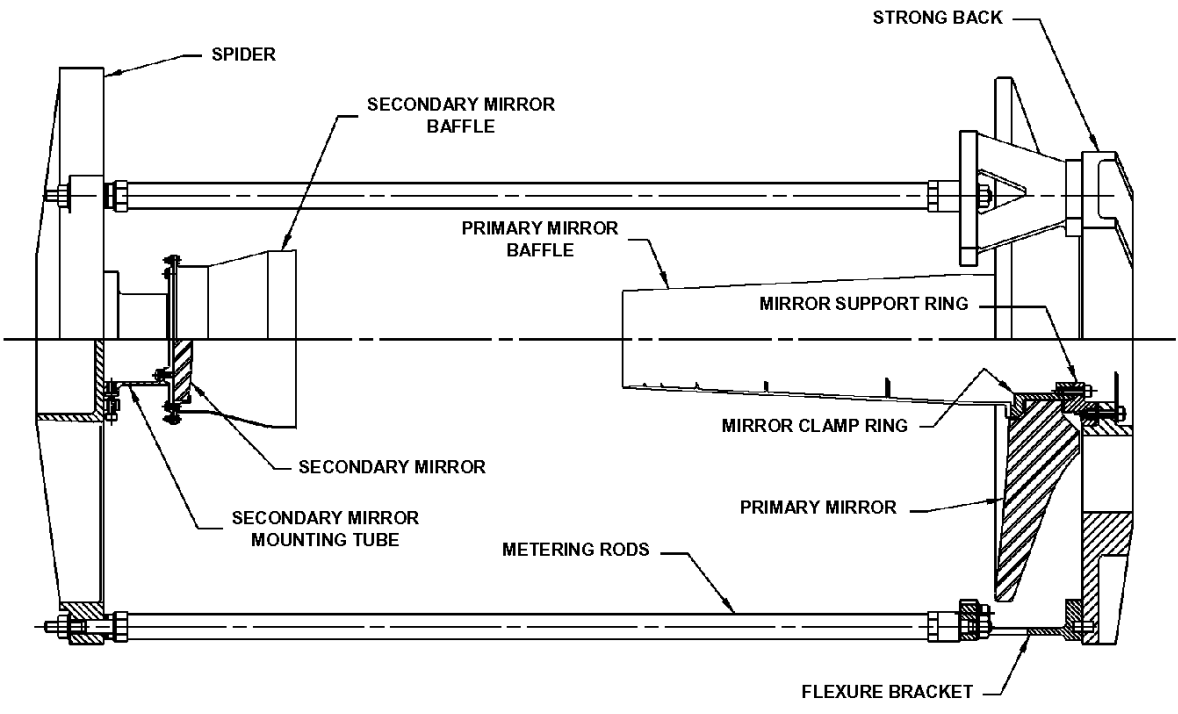

Figure 3

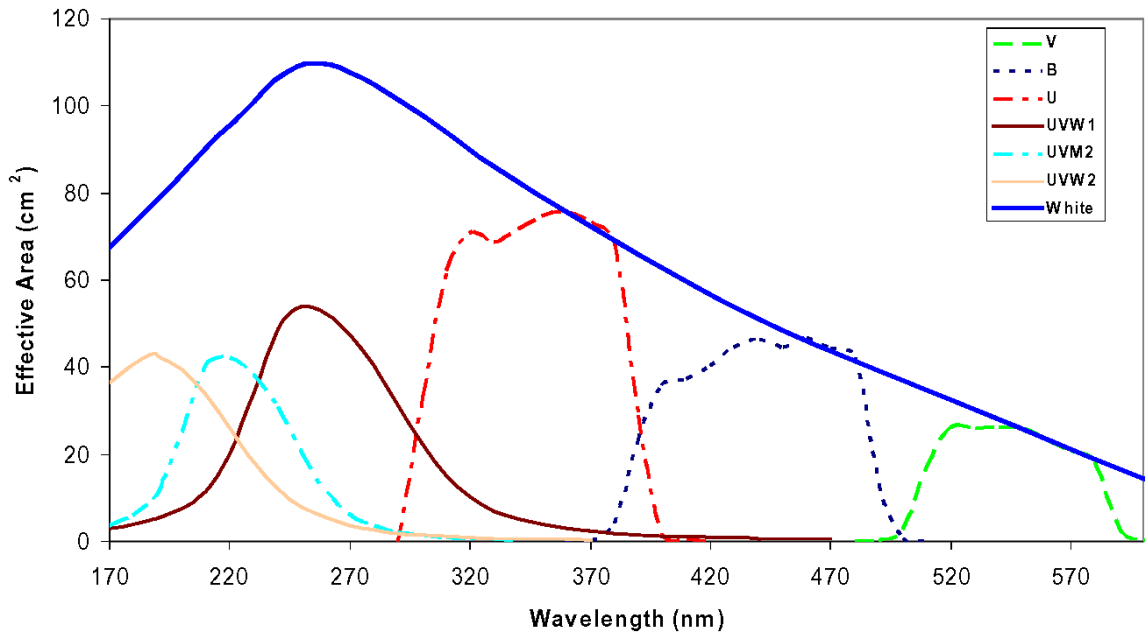

Figure 4

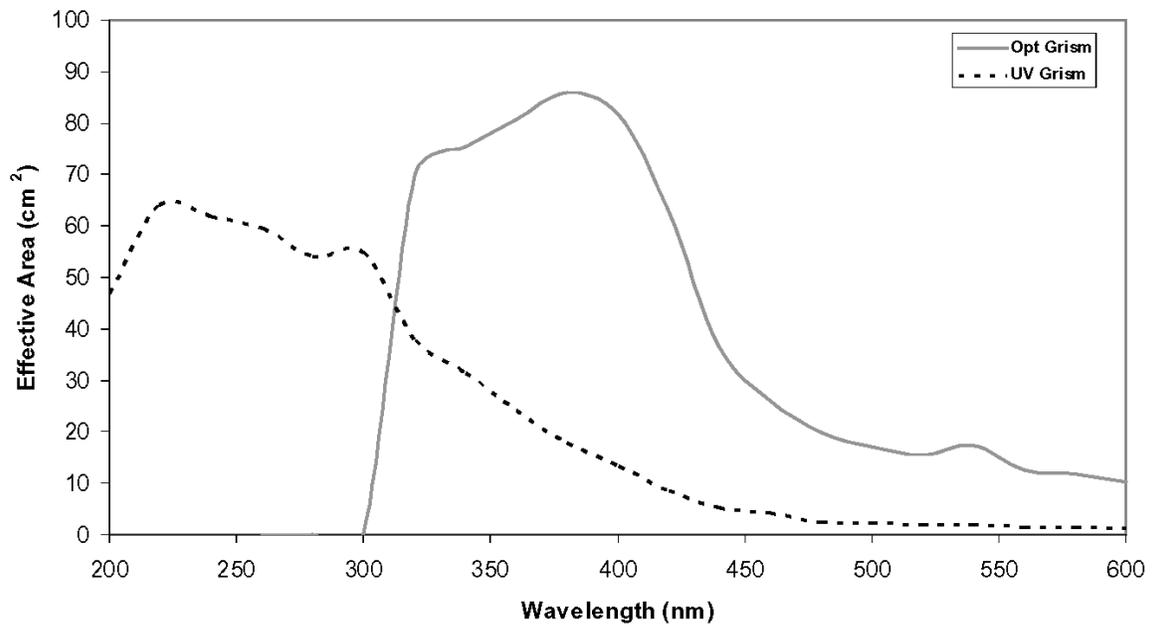

Figure 5

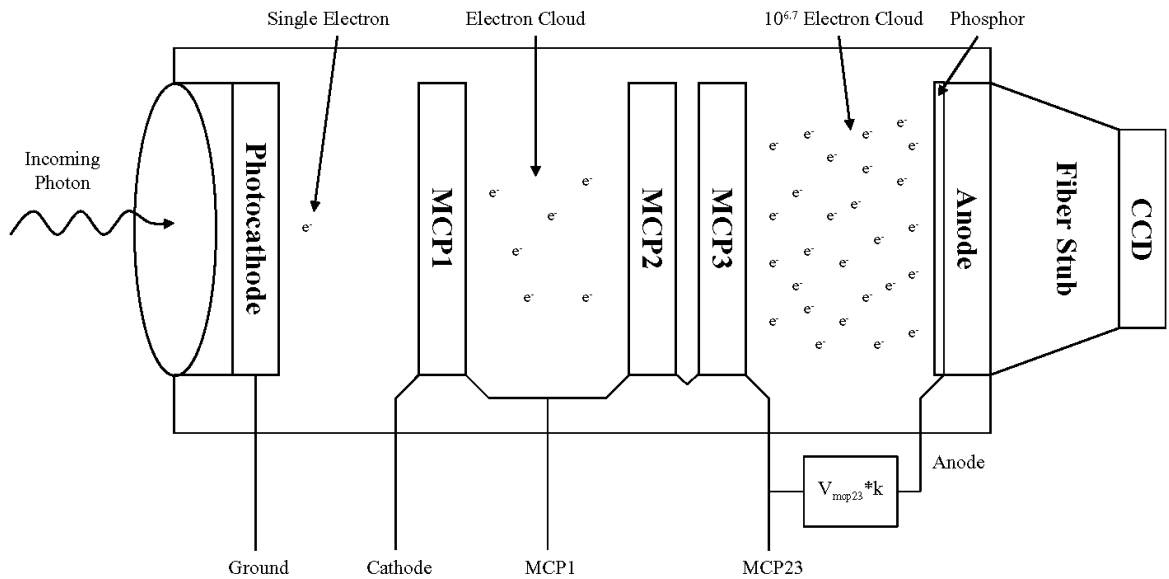

Figure 6

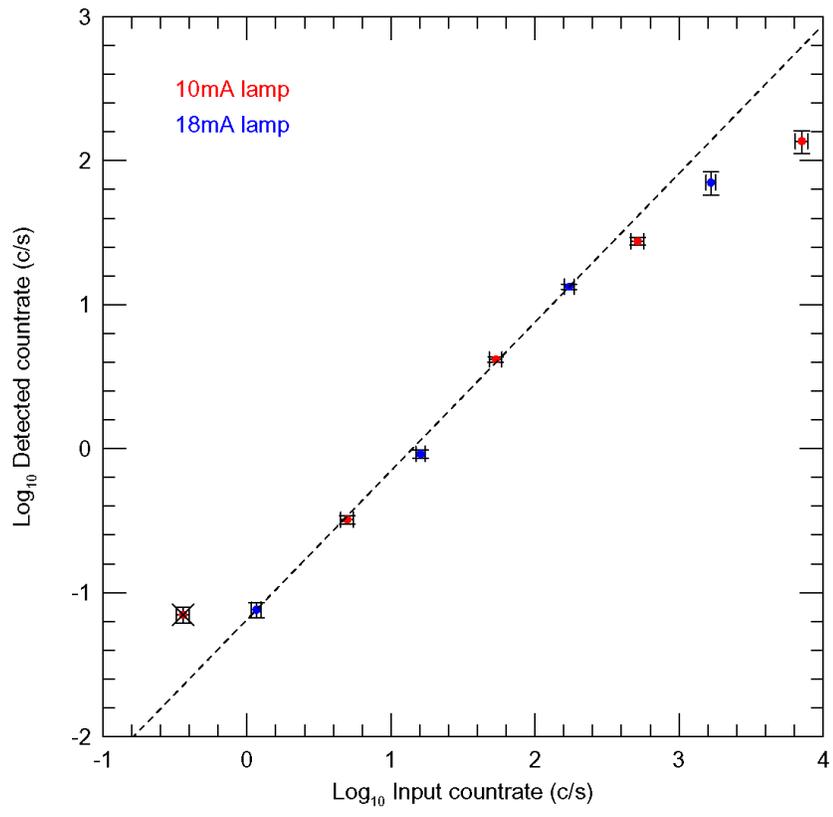

Figure 7

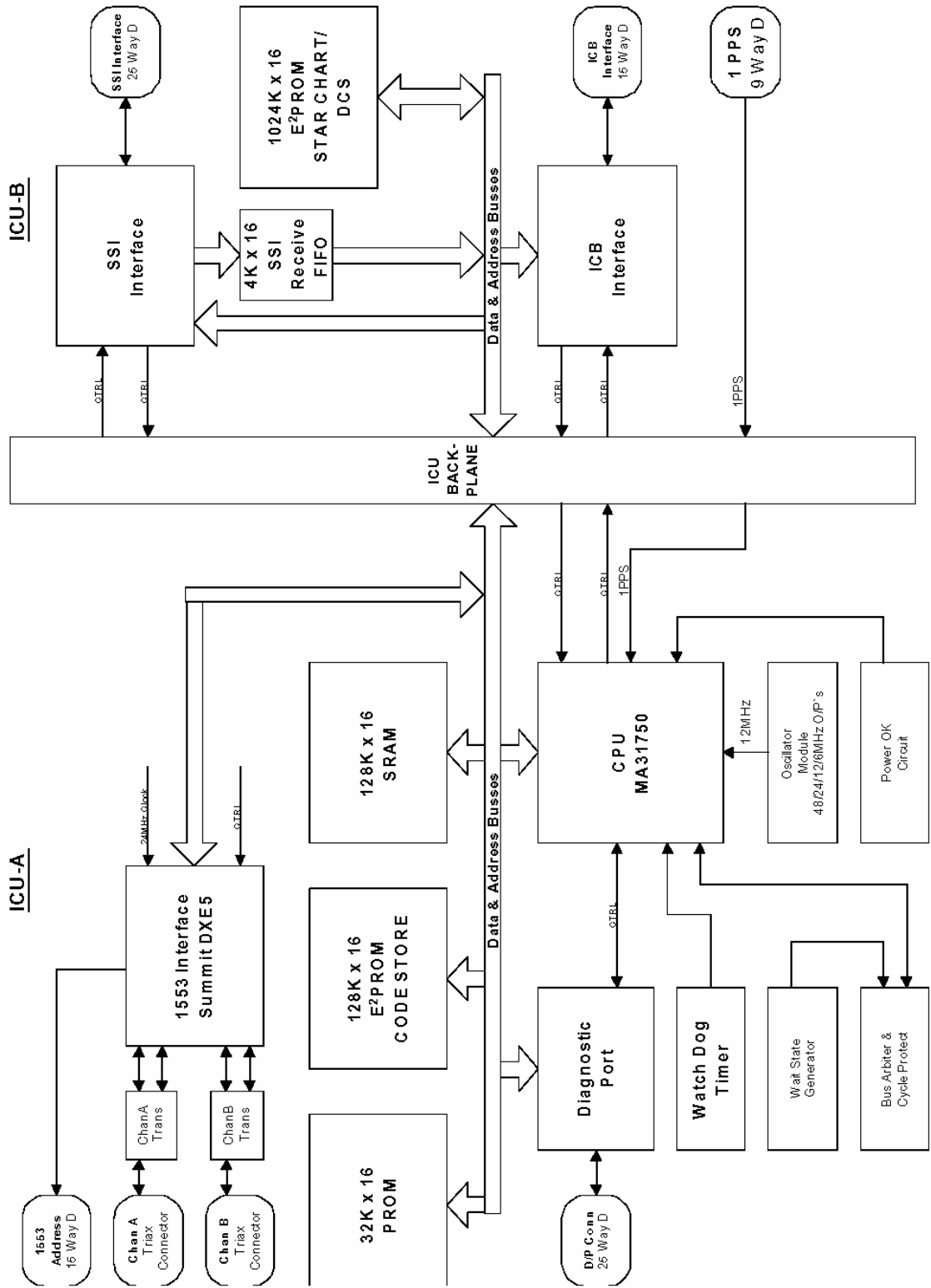

Figure 8

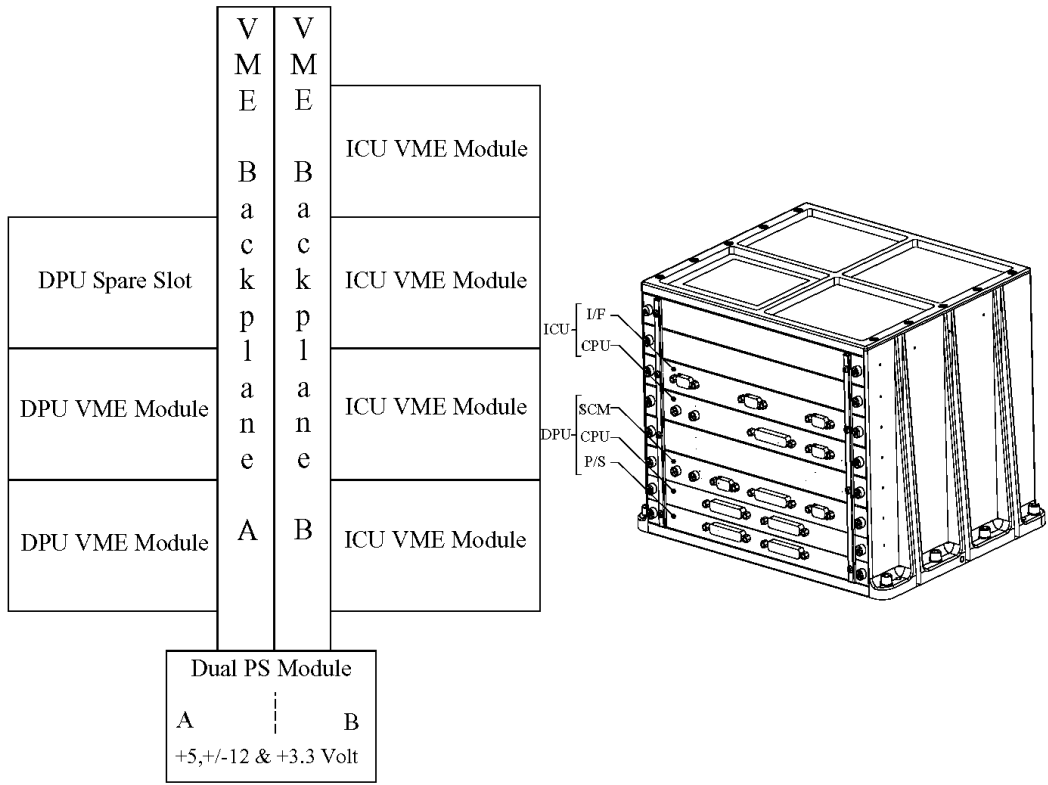

Figure 9

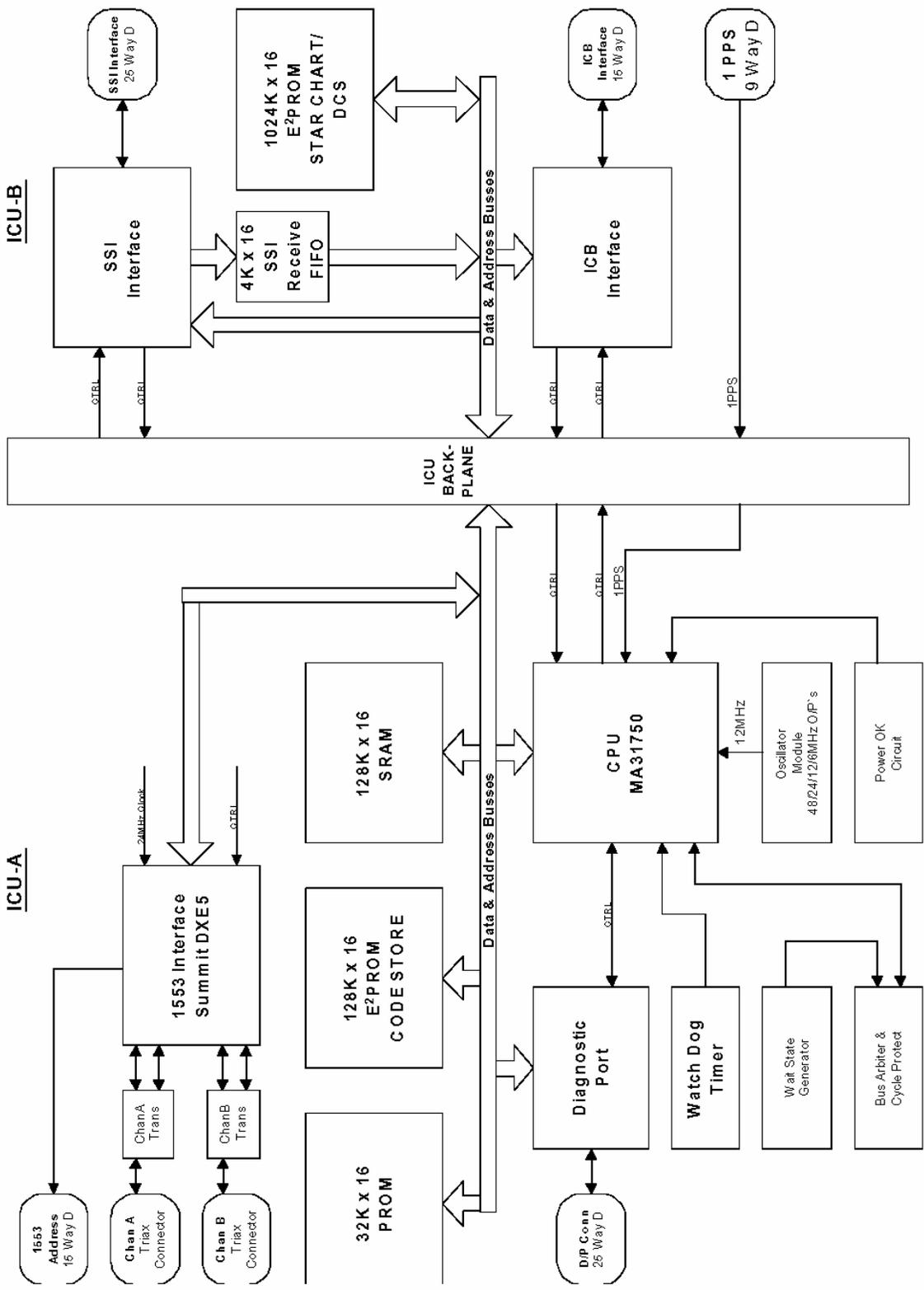

Figure 10

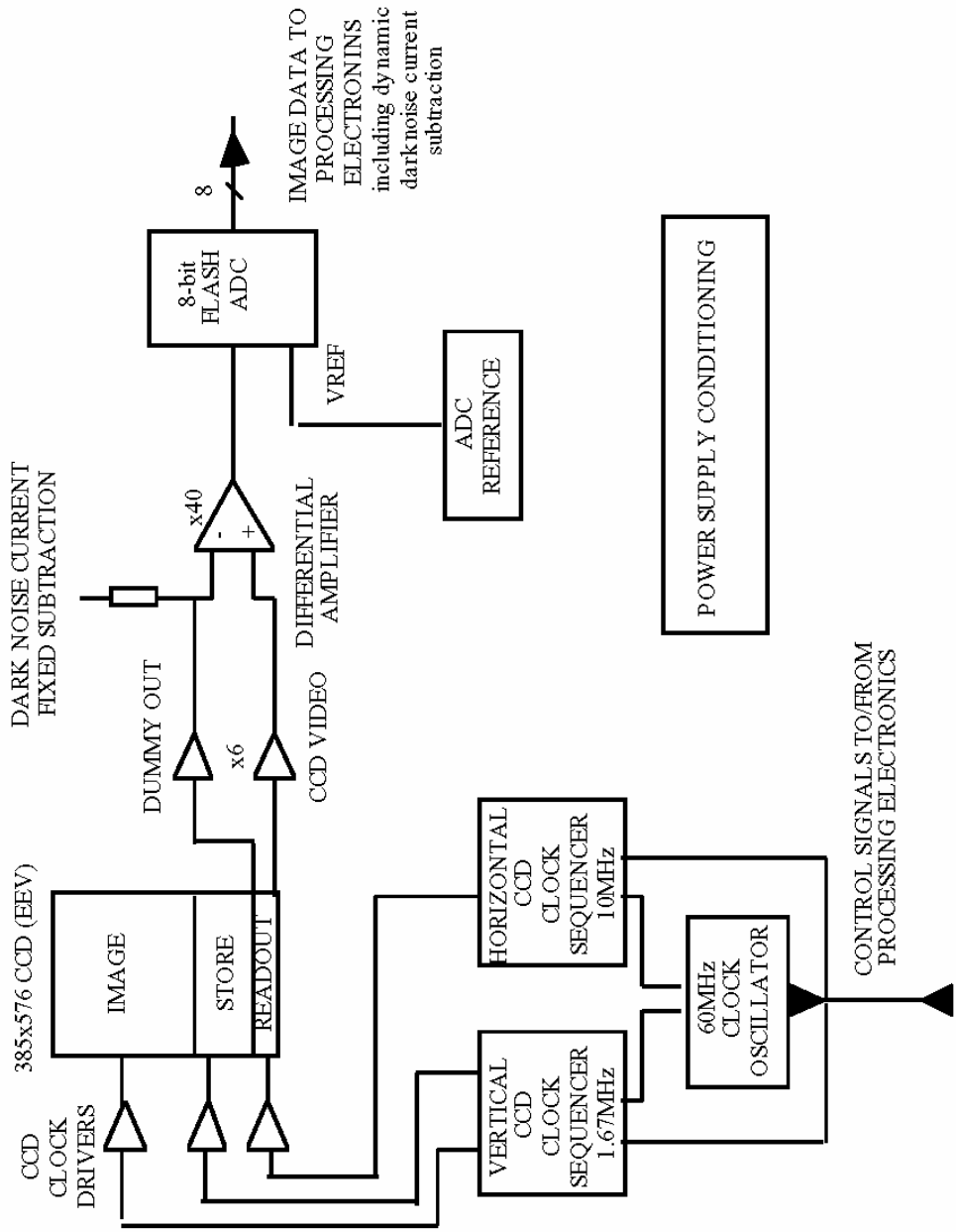

Figure 11

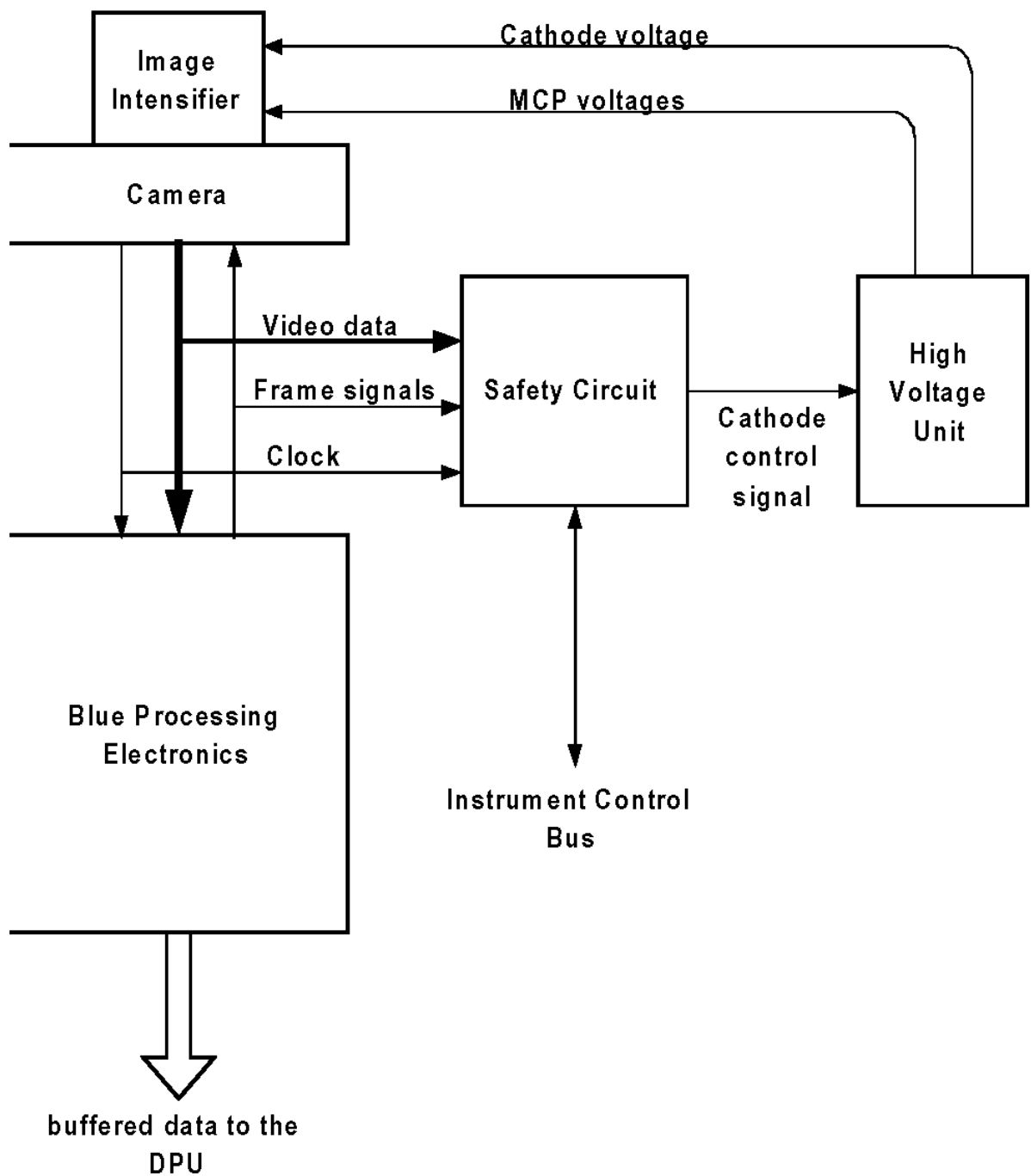

Figure 12

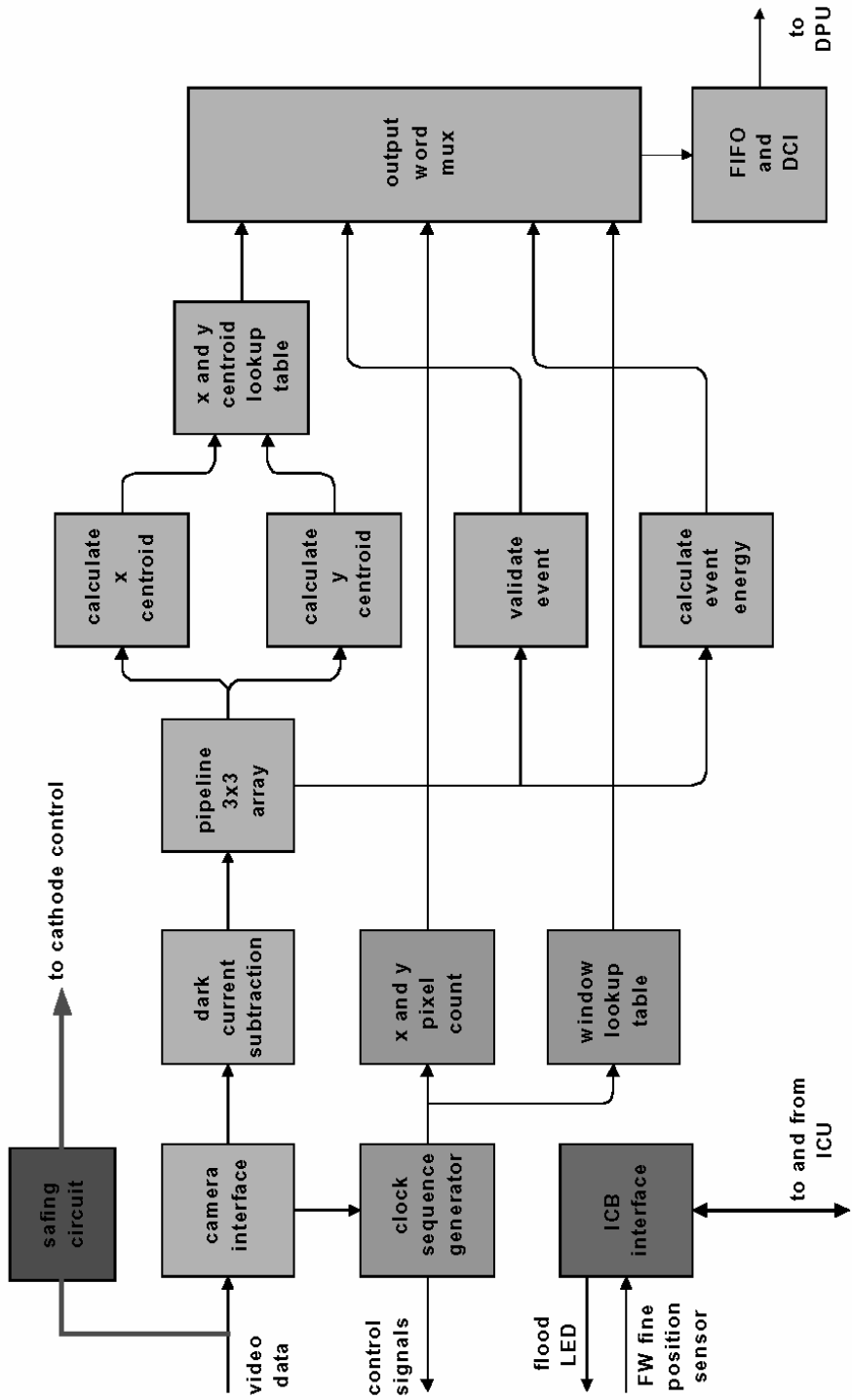

Figure 13

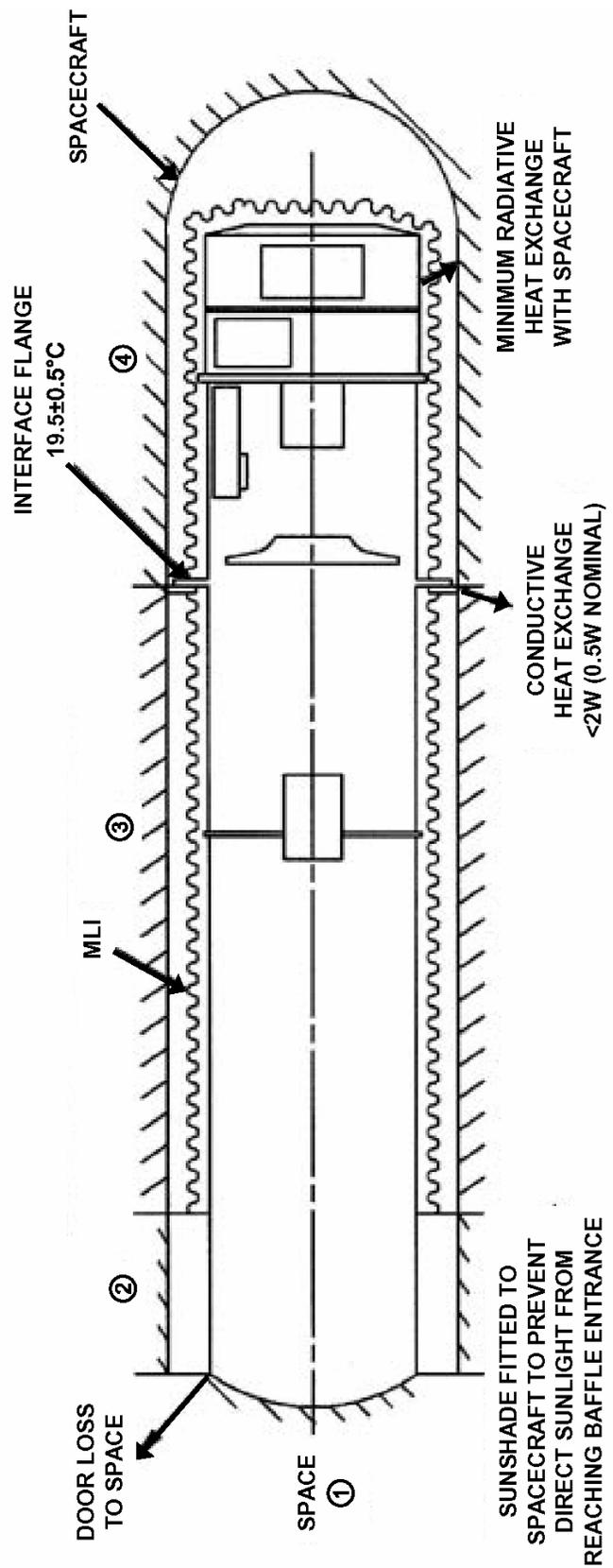

Figure 14

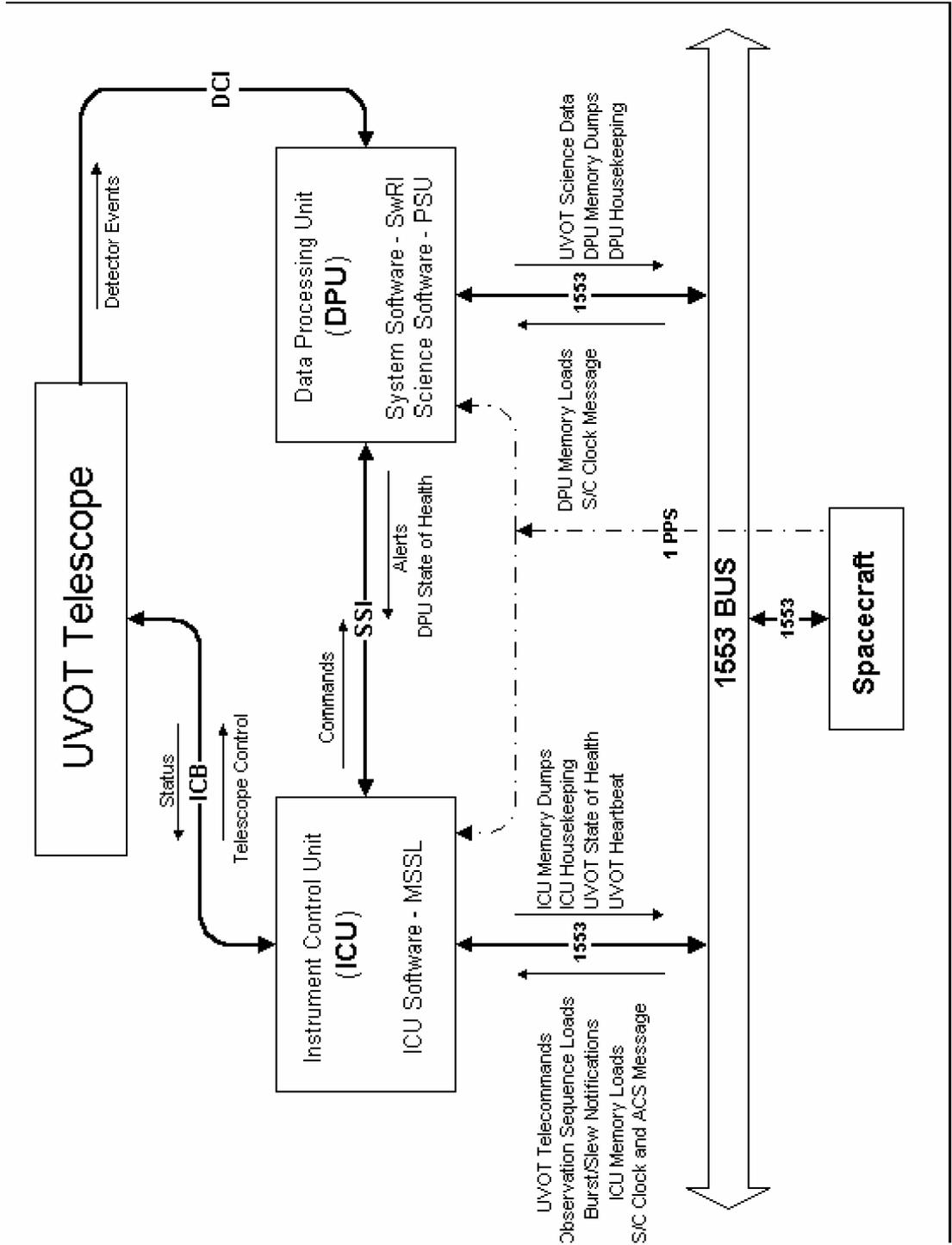

Figure 15

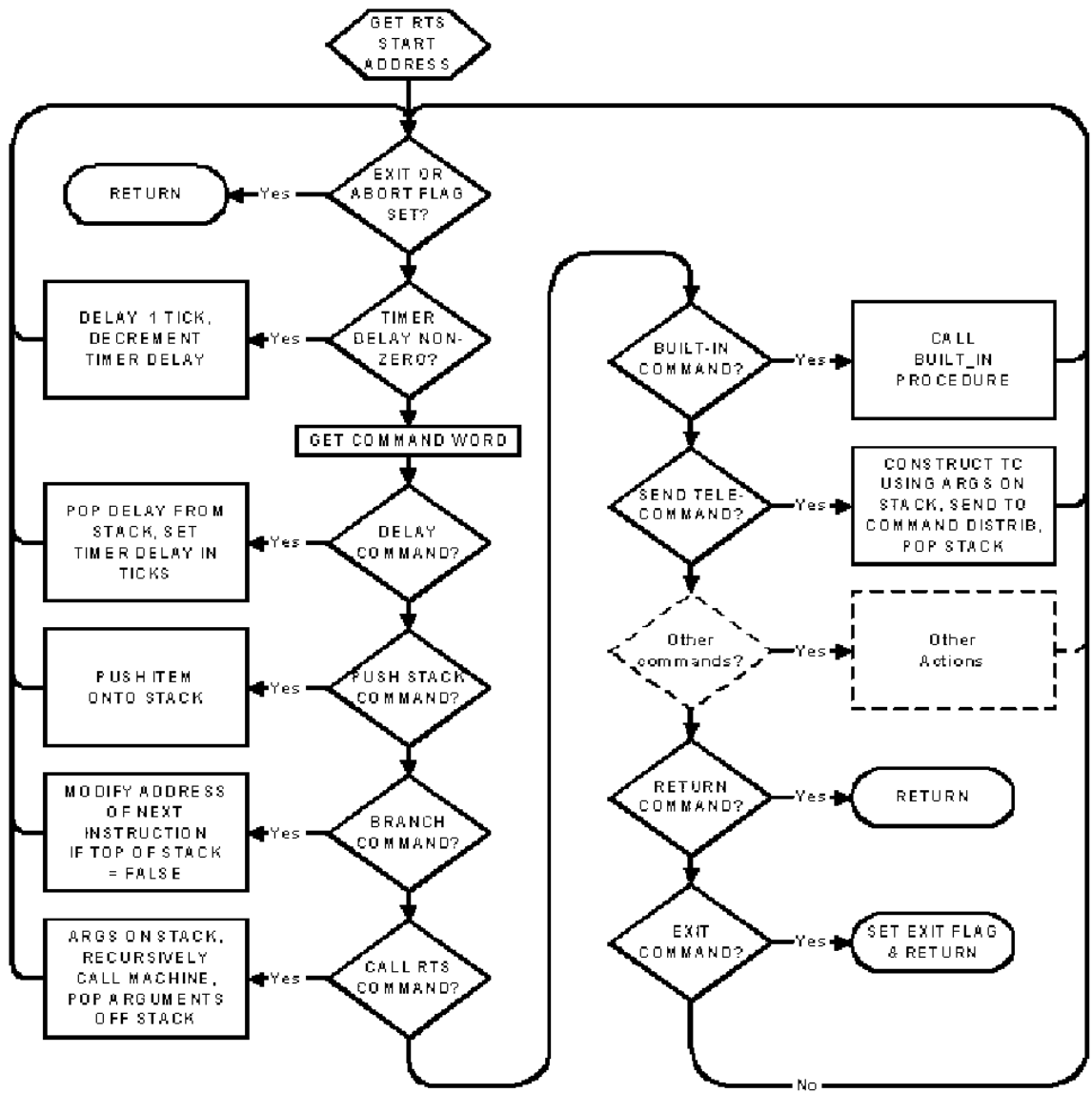

Figure 16

Figure 17

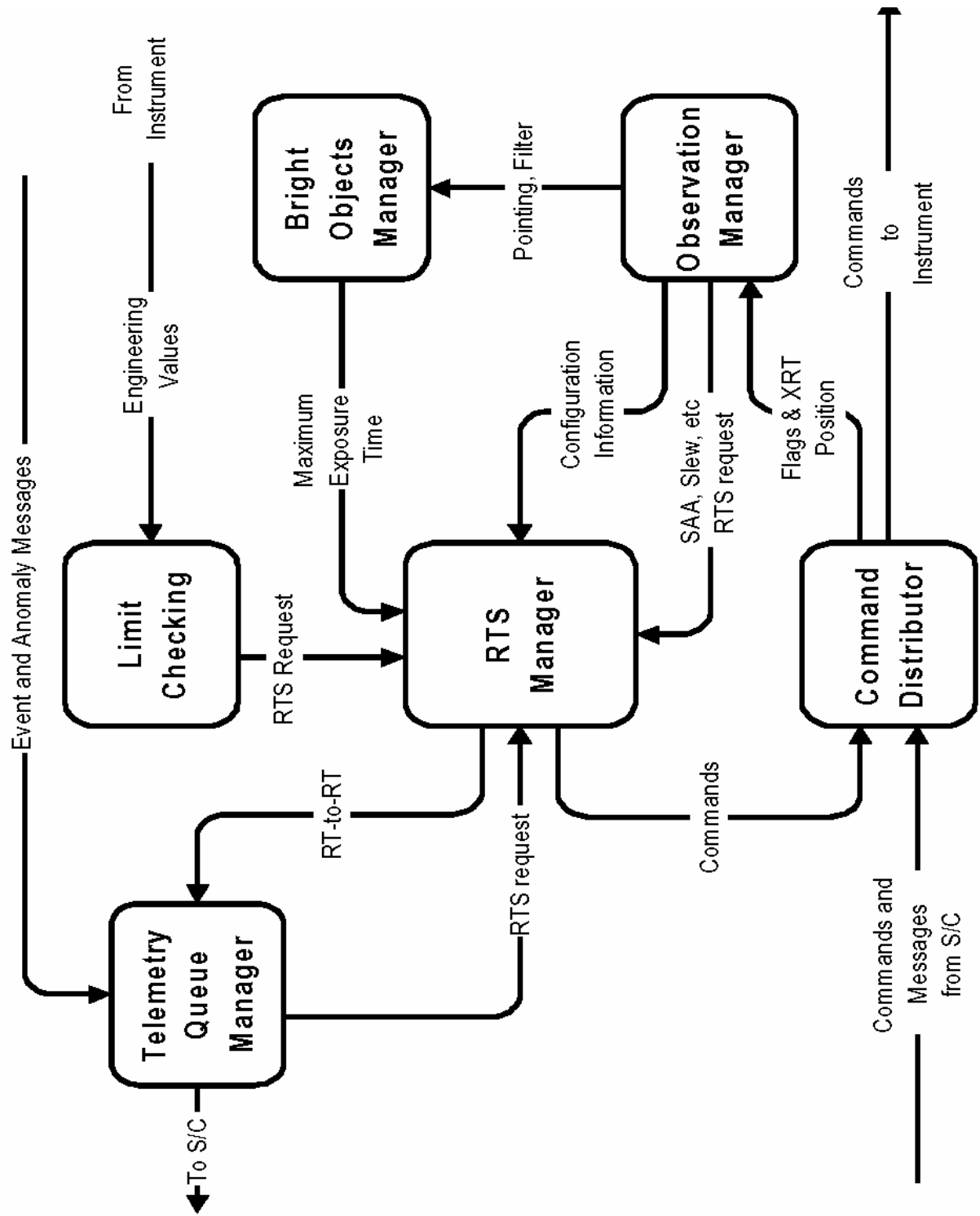

Figure 18

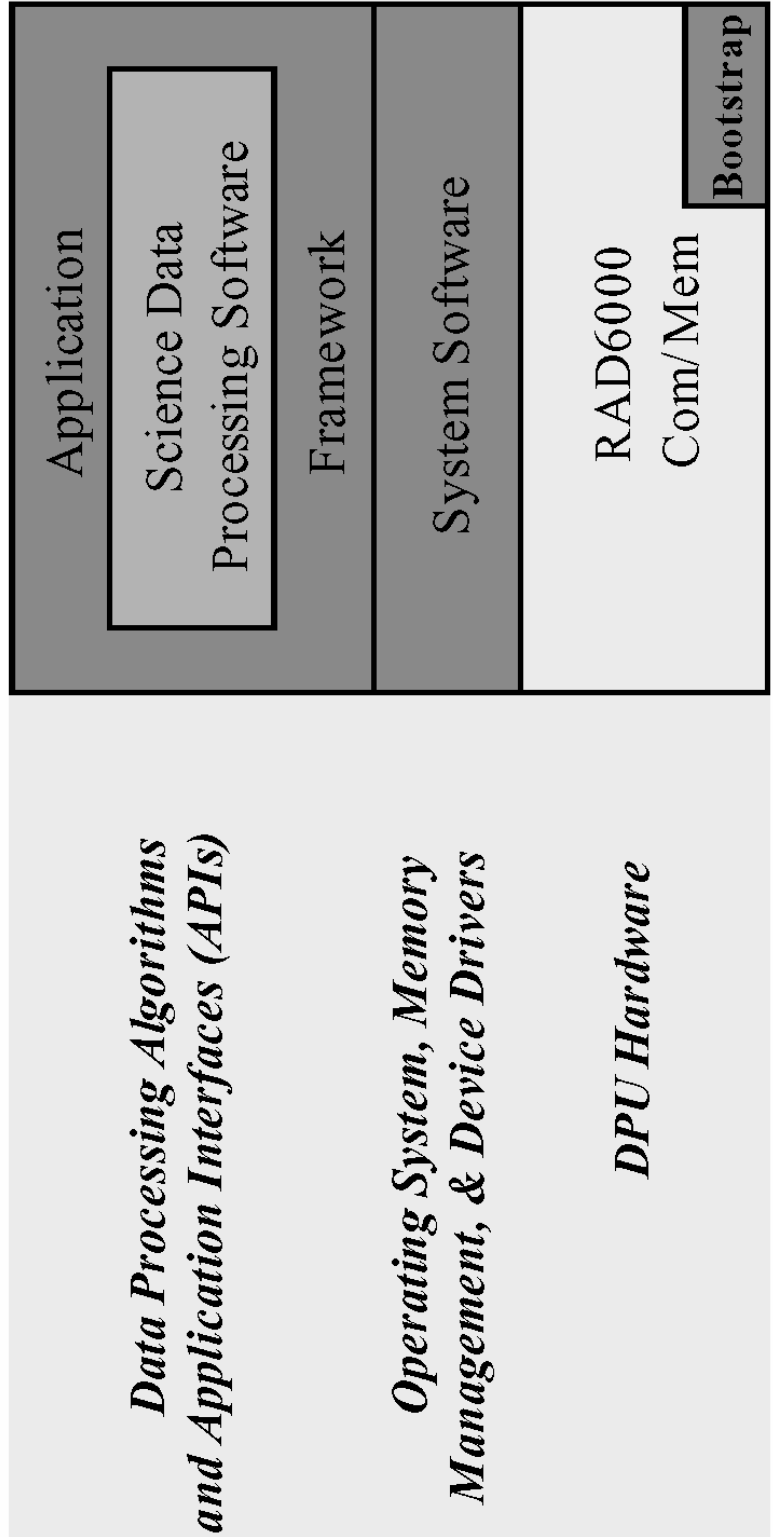

Figure 19

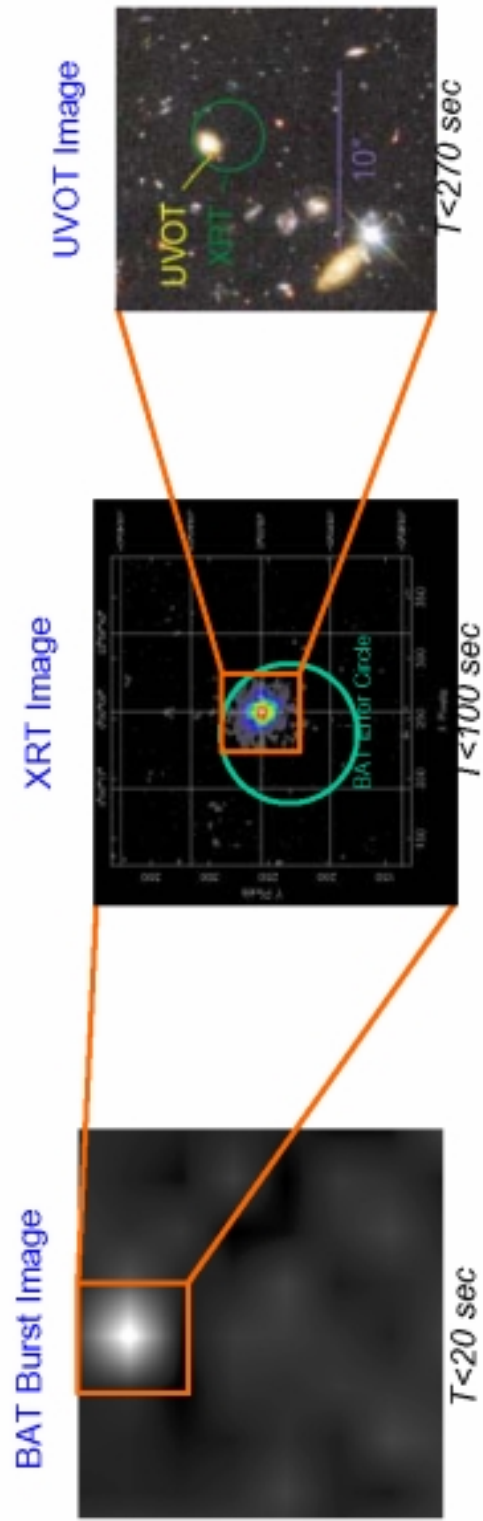

Figure 20

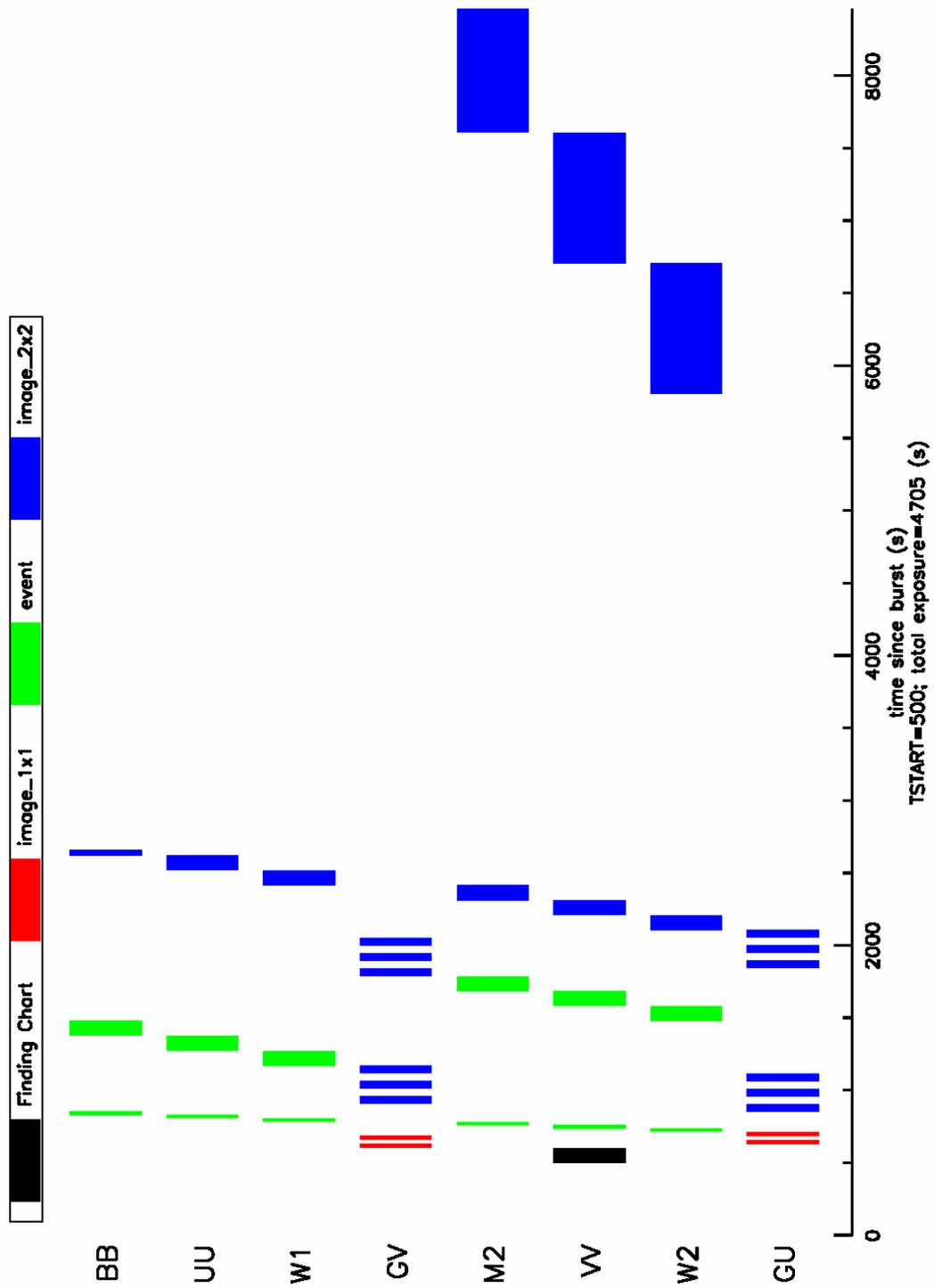

Figure 21

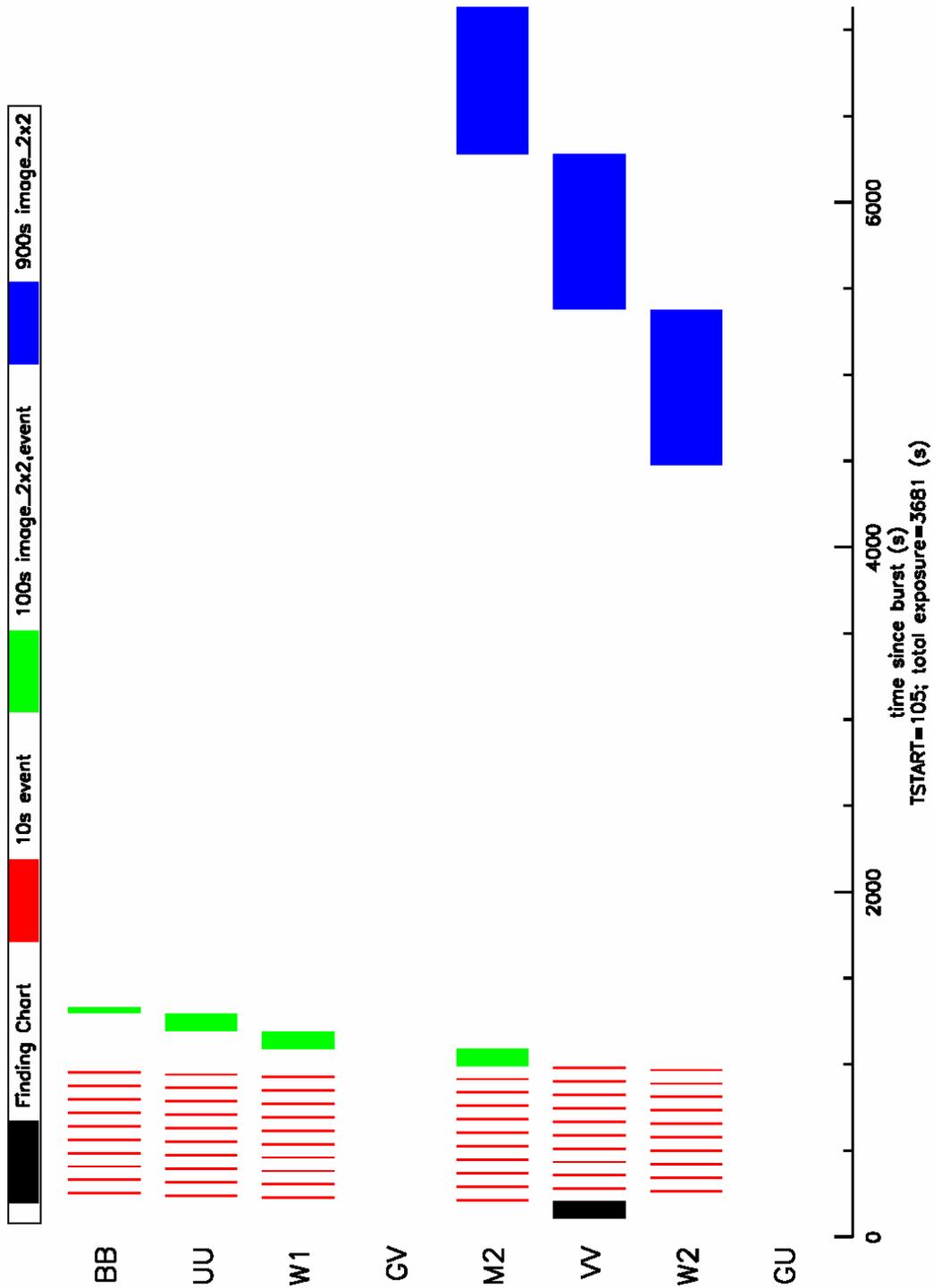

Figure 22

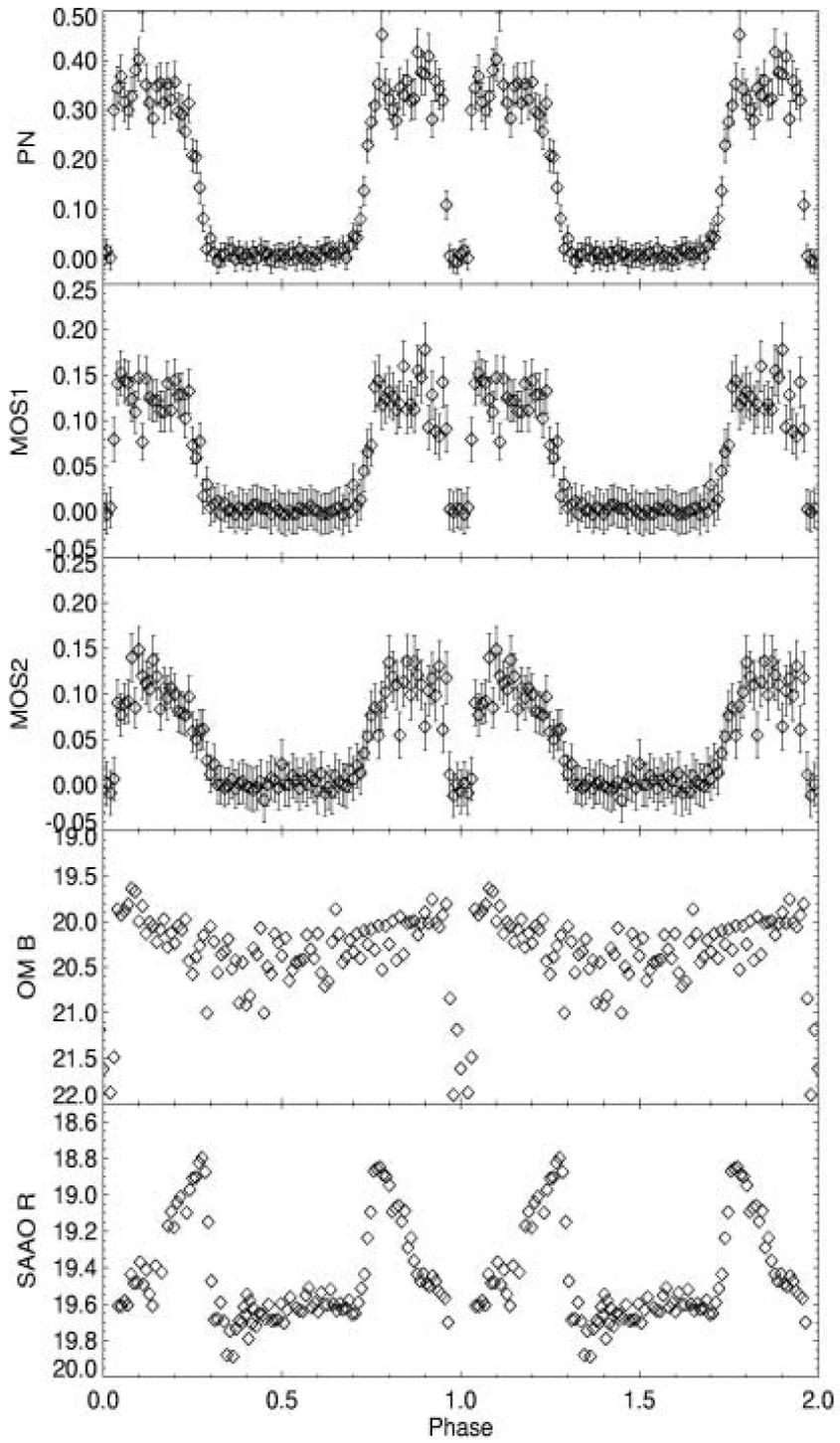

Figure 23

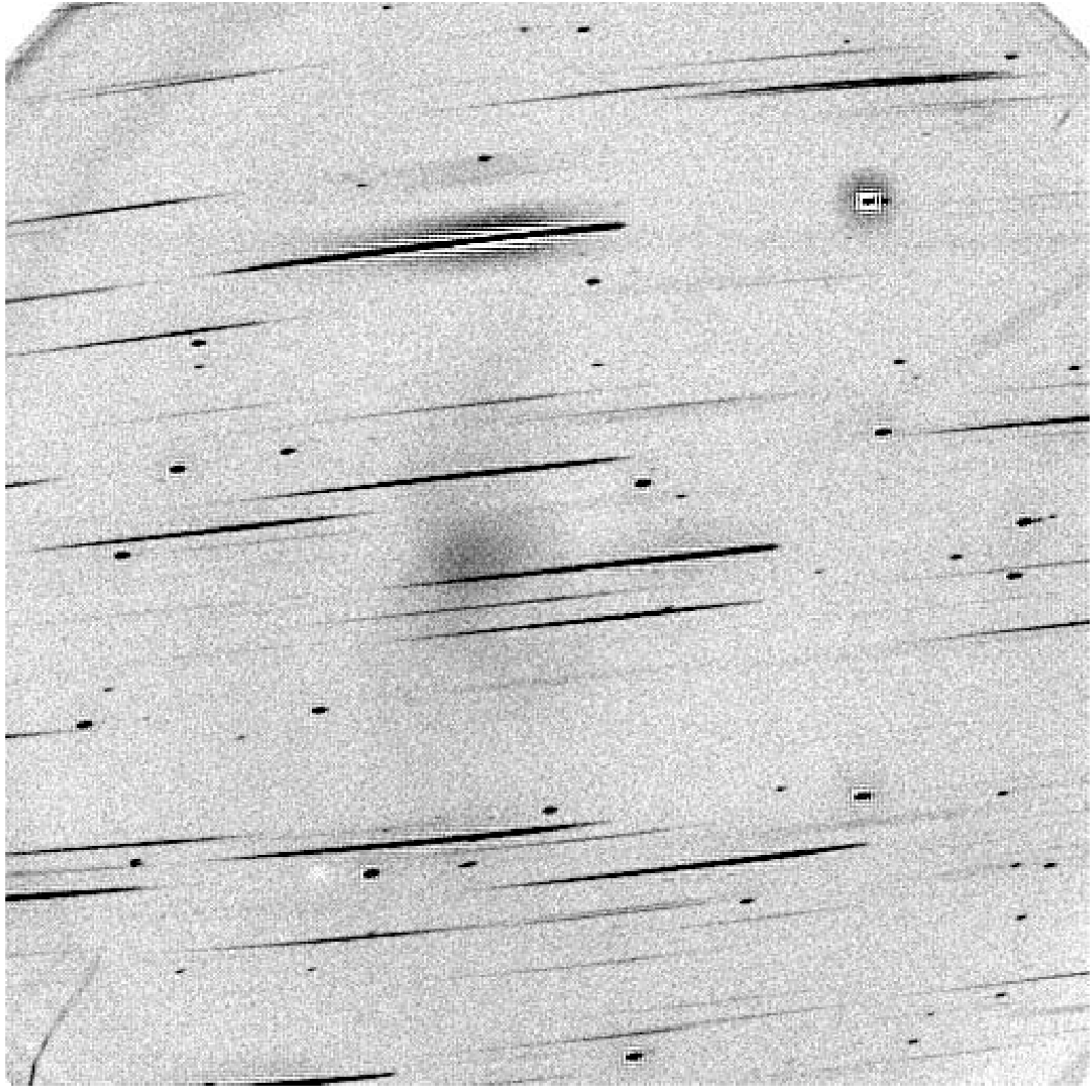

Figure 24

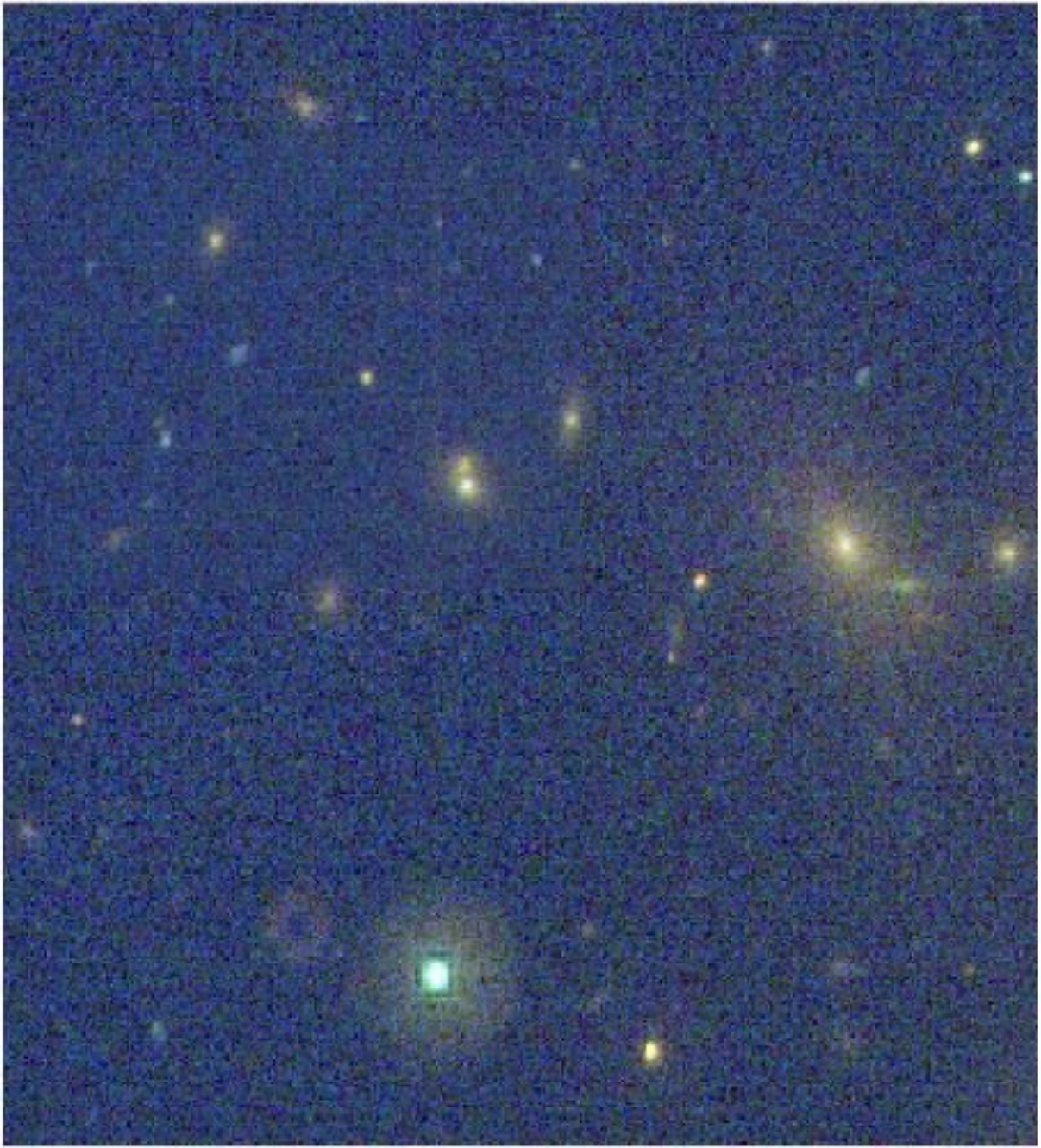

Figure 25

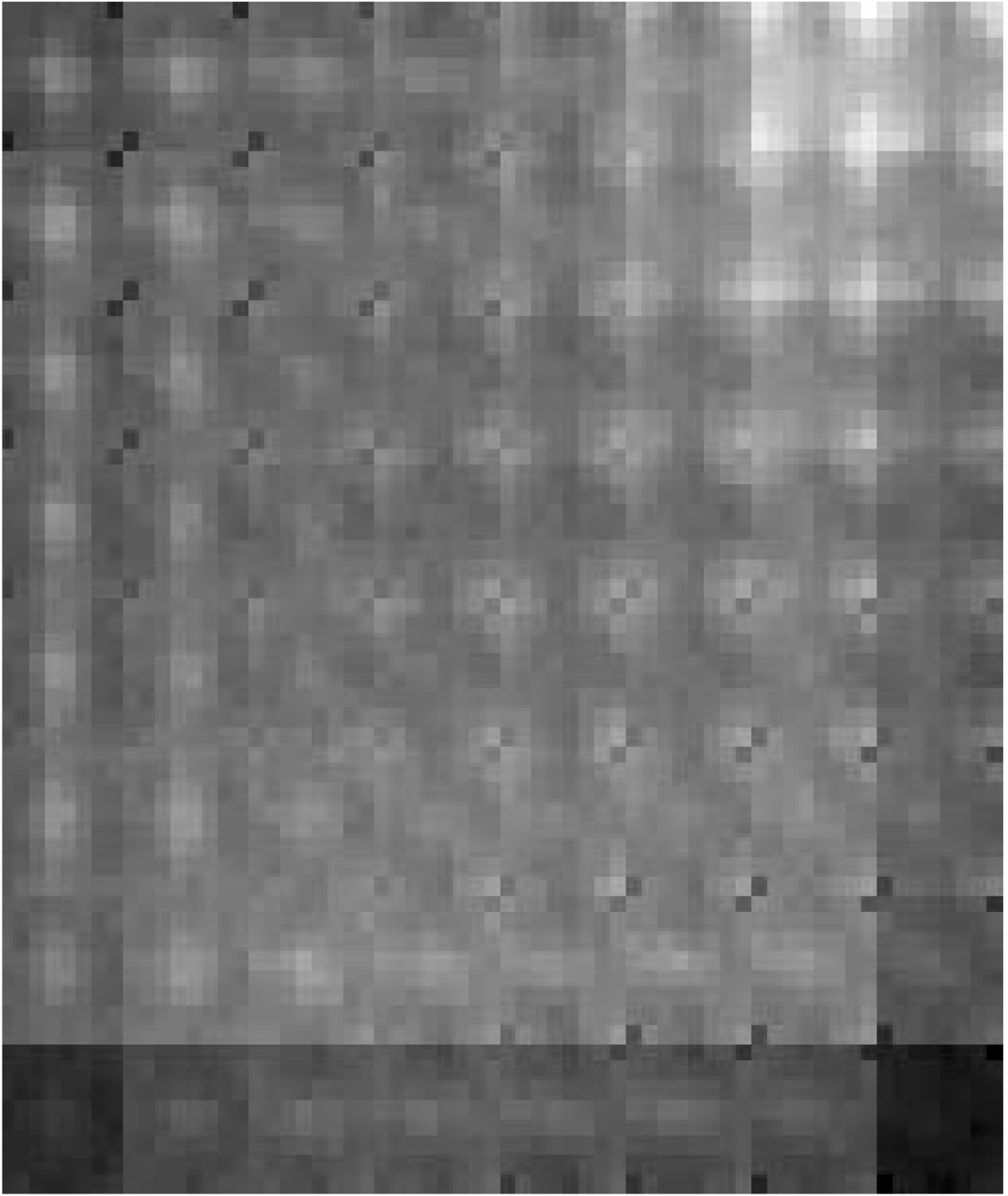

Figure 26

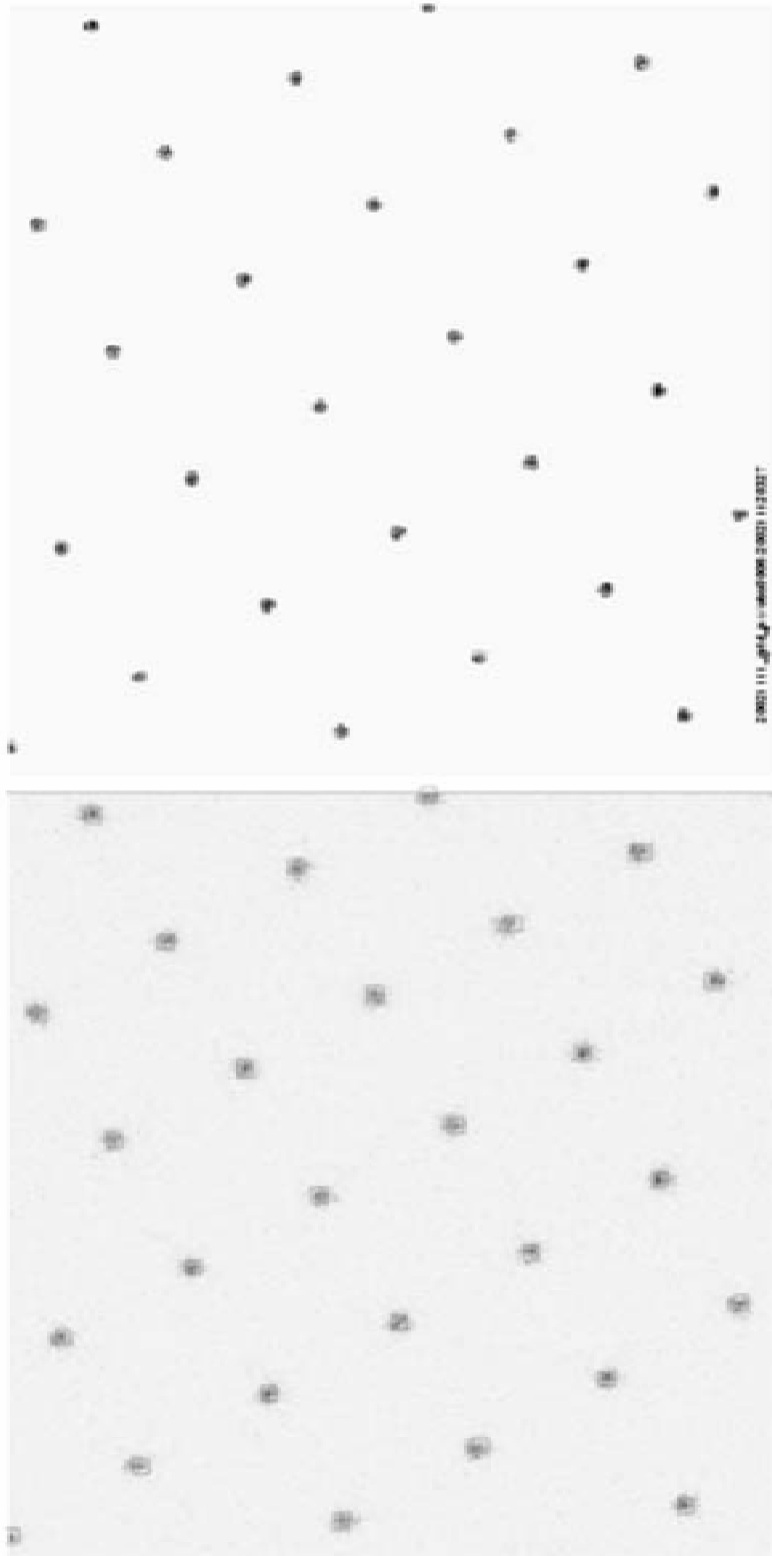

Figure 27

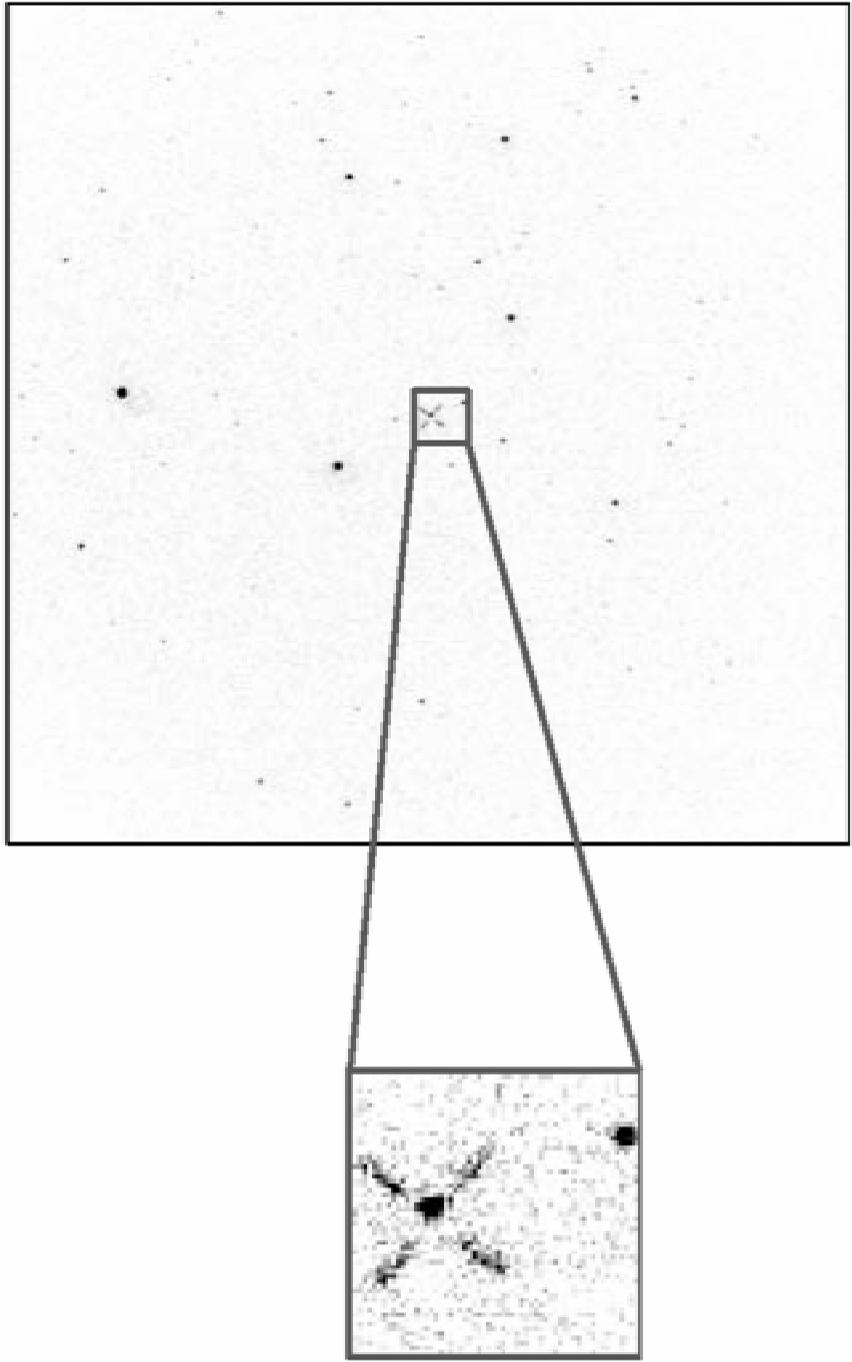

Figure 28